%% file: Catresubmission.tex
\documentclass[11pt]{article}
\usepackage{graphicx}
\usepackage{color}
\usepackage{float}
\usepackage{enumerate}
\usepackage{appendix}
\usepackage{empheq}
\usepackage{hyperref}
\usepackage{fancyhdr}
\definecolor{slidegreen}{rgb}{0,.5,0}
\definecolor{slidered}{rgb}{1,0,0}
\definecolor{slideblue}{rgb}{0,0,1}

\def\red#1{\textcolor{slidered}{#1}}
\def\blue#1{\textcolor{slideblue}{#1}}

\usepackage{amsmath}
\usepackage{amssymb}
\usepackage{amsmath}
\usepackage{amssymb}
\newcommand{\thedate}{\today}

\input{preamble.tex}

\newcommand{\mmp}{measurement problem}

\newcommand{\hv}{hidden variable}

\topmargin = - 0.5 cm \textheight = 23 cm \textwidth = 15 cm
\long\def\symbolfootnote[#1]#2{\begingroup%
\def\thefootnote{\fnsymbol{footnote}}\footnote[#1]{#2}\endgroup}
\begin{document} 
\pagenumbering{arabic} \setlength{\unitlength}{1cm}\cleardoublepage
\thispagestyle{empty}
\begin{center}
\Huge\textbf{A Flea on Schr\"{o}dinger's Cat}
\bigskip

\Large{N.P. (Klaas)  Landsman\symbolfootnote[1]{Institute for Mathematics, Astrophysics, and Particle Physics, Radboud University Nijmegen, \\
Heyendaalseweg 135, 6525 AJ  6525 AJ  Nijmegen,
The Netherlands.
Email: \texttt{landsman@math.ru.nl}}   \,  and Robin Reuvers\symbolfootnote[2]{Radboud Honours Academy, Faculty of Science, Radboud University Nijmegen. Current address: Pembroke College, Cambridge CB2 1RF, U.K.
Email: \texttt{robinreuvers@gmail.com}}
}
\bigskip

\thedate

\bigskip

\smallskip

 \textbf{Abstract}
 \end{center}
  We propose a technical reformulation of the measurement problem of quantum mechanics, which is 
 based on the postulate that the final state of a measurement is classical; this accords with experimental practice as well as with  Bohr's views. 
   Unlike the usual formulation (in which the post-measurement state is a unit vector in Hilbert space),
   our version actually opens the possibility of admitting  a purely technical solution within the confines of conventional quantum theory (as opposed to solutions that either modify this theory, or introduce unusual and controversial interpretative rules and/or ontologies). 
   
To that effect, we recall  a remarkable phenomenon in the theory of Schr\"{o}dinger operators
(discovered in 1981 by Jona-Lasinio, Martinelli, and Scoppola),
according to which the ground state of a symmetric double-well  Hamiltonian (which is paradigmatically of Schr\"{o}dinger's Cat type) becomes exponentially sensitive to tiny perturbations of the potential as
$\hbar\raw 0$. We show that this instability
emerges also from the textbook  {\sc wkb} approximation, extend it to time-dependent perturbations, and study the dynamical transition from the ground state of the double well  to the perturbed ground state (in which the cat is typically either dead or alive, depending on the details of the perturbation). 

Numerical simulations show that adiabatically arising perturbations may  (quite literally) \emph{cause} the collapse of the wave-function in the classical limit. Thus, at least in the context of a simple mathematical model, we combine the technical and conceptual virtues of decoherence (which fails to solve the \mmp\ but launches the key idea that perturbations may come from the environment) with those of
 dynamical collapse models \`{a} la {\sc grw} (which do solve the \mmp\ but are ad hoc), without sharing their drawbacks:   single measurement outcomes are obtained (instead of merely  diagonal reduced density matrices), and
no modification of quantum mechanics is needed.
\bigskip
 \begin{small}
 \begin{center} \textbf{Motto}\end{center}
 \begin{quote}
`Another secondary readership is made up of those philosophers and physicists who---again like myself---are puzzled by so-called foundational issues: what the strange quantum formalism implies about the nature of the world it so accurately describes. (\ldots) My presentation is suffused with a perspective on the quantum theory that is very close to the venerable but recently much reviled Copenhagen interpretation. Those with a taste for such things may be startled to see how well quantum computation resonates with the Copenhagen point of view. Indeed, it had been my plan to call this book \emph{Copenhagen Computation} until the excellent people at Cambridge University Press and my computer-scientist friends persuaded  me that virtually no members of my primary readership would then have any idea what it was about.'

\hfill David Mermin,  \emph{Quantum Computer Science: An Introduction} (Preface)
\end{quote}
\end{small}
\newpage
\section{Introduction}
Citizens of many democratic countries know the phenomenon of a ``hung parliament'', in which two major political parties have a large number of seats each, but are short of a majority separately and mutually exclude each other as potential coalition partners. In that case, tiny parties with just a few seats can tip the balance to the left or to the right and hence, quite undemocratically,  acquire an importance far exceeding their relative size. 

An analogous phenomenon in quantum mechanics was  discovered in 1981 by Jona-Lasinio, Martinelli, and Scoppola 
\cite{JLMS1,JLMS2} (see also Section \ref{FE} below). Here,  the ground state of (say) a symmetric double-well  Hamiltonian becomes 
exponentially sensitive (in $1/\hbar$)  to tiny perturbations of the potential as $\hbar\raw 0$. 
In particular, whereas the ground state of the unperturbed Hamiltonian has two pronounced and well-separated peaks, the ground state of the perturbed Hamiltonian typically features one of those peaks only and hence may be said to have ``collapsed''. We will  call such a perturbation a ``flea'' \cite{Simon4}.

As we will explain and make precise in Section \ref{MP}, this  phenomenon acquires acute relevance for the \mmp\ as soon as 
one accepts just two postulates:\footnote{We regard these as Wittgensteinian ``hinge propositions'' \cite{OC}, on which modern physics is based.}
\begin{enumerate}
\item \emph{A measurement yields a classical snapshot (or ``readout'') of a quantum state.}\footnote{This is sometimes called Bohr's \emph{doctrine of classical concepts} \cite{Scheibe},
 which accurately describes experimental practice!
 See also \cite{handbook} for a detailed analysis. }
 \item \emph{The fundamental nature of quantum theory.} This implies in particular that measurement devices (like anything else) are ultimately quantum mechanical in nature.\footnote{For our work it is a moot point whether Bohr endorsed this second point as well; it is hard to say.}
\end{enumerate}
By the first clause of this conjunction, the post-measurement state of the pertinent apparatus should be a classical state, whilst by the second, it has to be  the classical limit of some quantum state.
 After coupling to some microscopic object, the latter state might evolve into a superposition \`{a} la Schr\"{o}dinger's Cat, and this is what causes the \mmp.  But it is exactly in the classical limit that the sensitivity of the wave-function to the flea arises! Thus the correct formulation of the \mmp, viz.\   as a problem concerning classical limits of quantum states, already contains the seed of its solution. 
 
In Section \ref{FE} we review \emph{static} aspects of ``flea'' instability (using the two-level system as a pedagogical example), in that the perturbations are taken to be time-independent, and the perturbed ground state is just studied for its own sake. This review is backed up by a new, more technical analysis based  on the  {\sc wkb} approximation familiar from the textbooks,\footnote{The original expositions \cite{CDS2,GGJL,HS,JLMS1,JLMS2,Simon4} might be hard to follow for non-mathematicians (see, however, \cite{CRT}).  Applications in chemistry and solid-state physics may be found in \cite{ClJo86,JLPT,MS1}.} which we delegate to the appendix in order not to interrupt our story. 
In order to trace the fate of Schr\"{o}dinger's Cat as a dynamical process, we need to take up the \emph{dynamical} study of the 
instability. In other words, the perturbation should be made time-dependent, and in addition the transition from the unperturbed ``Schr\"{o}dinger Cat'' ground state to the perturbed ground state (in which the cat is either dead or alive) should be followed in time. This will be done (mostly numerically) in Section \ref{Dynamical}, showing that the analysis of the \mmp\ given here in a toy example  (including a derivation of the Born rule) has a fighting chance of eventually being able to solve the problem.

In the closing Discussion section, we explore: possible generalizations of our approach, 
connections  with  symmetry breaking and phase transitions, the role of quantum metastability, and finally the role of determinism and locality (vis-\`{a}-vis Bell's Theorem etc.).
 \section{Rethinking the \mmp}\label{MP}
 \subsection{Historical overview}\label{histo}
 Roughly speaking, the measurement  problem consists in the fact that the Schr\"{o}dinger equation of quantum mechanics generically fails to predict that measurements have outcomes. Instead, it apparently predicts (empirically) unacceptable ``superpositions'' thereof.  Slightly more technically, the problem is usually formulated in approximately the following way
(see e.g.\ the excellent presentations in \cite{BG,Bub,BLM} for more detail). Suppose that one measures some observable $O$ pertaining to 
 a microscopic system $S$, in such a way that if $S$ is in an eigenstate  $\Phi_i$ of $O$, then the associated macroscopic apparatus $A$ is in state $\Om_i$. It is important to note that in this description  (pure) states are seen as unit vectors in some Hilbert space, as usual in quantum mechanics.
Now, although there is hardly any  problem with the existence of a \emph{microscopic} superposition $\Phi= \sum_i c_i \Phi_i$ of $S$ (where $\sum_i |c_i|^2=1$), the linearity of the  Schr\"{o}dinger equation implies that it would bring $S+A$ into a similar superposition  $\Ps=\sum_i c_i \Psi_i$, $\Psi_i\equiv\Phi_i\otimes \Om_i$, which,  \emph{if macroscopic},\footnote{This notion needs to be quantified, of course \cite{BG,BLM}. See also \cite{handbook} and the present paper.} is never seen in nature. Instead,  as a matter of fact one always observes \emph{one} of the states $\Om_i$ (or so it is claimed). 

This problem was immediately recognized by the founders of quantum theory.  In response, in 1926, Born (generalizing earlier ideas of  Bohr and Einstein on light emission by atoms) stated that quantum theory indeed did not predict individual outcomes, but merely computed their probabilities (according to the formula now named after him) \cite{Born,LanBorn}. In 
1927, Heisenberg (again in the wake of Bohr's  electronic ``quantum jumps'') proposed the ``collapse of the wave-packet'' \cite{Hei1927}, which (in the above language) implies that during the course of the measurement, the  state $\Ps$  miraculously ``jumps'' to one of the states $\Ps_i$.  Bohr immediately endorsed this idea, ordered that such quantum jumps ought not to be analyzed any further, and claimed that they were the source of irreducible randomness in physics. During the period 1927--1935 (with an aftermath running until 1949), Bohr famously defended these ideas against a highly critical  Einstein \cite{Bohr1949}, as did Born \cite{BEB}. Adding considerable conceptual and mathematical precision, von Neumann gave an account of the measurement  problem  in his book from 1932 \cite{vN32}, which has formed the basis for most discussions of the  issue ever since. The early period was closed in 1935, the year in which 
 Schr\"{o}dinger  (following a correspondence with Einstein \cite{FineSG}) published a penetrating analysis, including the metaphor around the cat later named after him \cite{Sch1935}.
 
The introduction of the collapse process (and the associated Born probabilities) has turned out to be an incredibly successful move, on which practically all empirical successes of quantum theory are based. This may well be the reason why few physicists are bothered by the measurement problem: 
 the underlying slick manoeuvre of ad hoc collapse seems an acceptable price for
these successes, unprecedented in science as they are. But for those in the foundations of physics, there is no doubt that this is a pseudo-solution.
Consequently, despite arguments claiming to prove their non-existence \cite{Brown,BLM,Fine}, 
many solutions to the measurement problem have been proposed. Among those, it is fair to say that at least the quasi-philosophical solutions have failed to 
 convince the scientific community at large.\footnote{This is true in particular for  the Many-Worlds (aka ``Everett'') Interpretation \cite{MWI}, or the Modal Interpretation of quantum mechanics \cite{Modal}, in which radical changes are proposed in the ontology and/or the usual interpretative rules of the theory, without clarifying in any way what is really going on during measurements (a question one indeed is not supposed to ask, according to received wisdom). Bohmian mechanics (as a modern incarnation of de Broglie's pilot-wave theory) does a better job here  \cite{CushingFine,DGZ,DT}, but its narrow applicability (at least in its current form), focusing as it does on position as the only physical observable, makes it unattractive to many (including the authors). See some kinship in \S\ref{Bell}, however.}

Within the realm of technical approaches to the measurement problem, one may distinguish between those proposals that do and those that do not modify quantum theory. Among the latter, 
the main effort so far has been towards attempts to eliminate interference terms (i.e., between  states like the $\Psi_i$ above), sometimes accompanied by the (implicit or explicit) suggestion that their removal would actually solve the problem.\footnote{This applies, for example, to the famous paper by Danieri,  Loinger, and Prosperi \cite{DLP}, to
early papers on decoherence \cite{Zurek1}, and to much of  the mathematical physics literature on the \mmp, including the work of the senior author \cite{EWW,Hepp,Landsman1991,KlaasObs,Sewellrecent}. We now regard such papers as mathematically interesting but conceptually misguided, at least on this point. On the other hand,  it is to the credit of especially the Swiss school that it drew attention to the
idea that measurement involves limiting procedures, so that solutions of the \mmp\ should at least \emph{incorporate} the appropriate limits. } 
 
 Such attempts come in (at least) two kinds. 
 In the wake of the Swiss school \cite{EWW, Hepp}, mathematical physicists typically use  the formalism of \emph{superselection rules} (see \cite{Landsman1991,KlaasObs} for reviews and \cite{Sewellrecent} for recent work in this direction), whereas theoretical physicists tend to exploit \emph{decoherence} (see \cite{Joos,Schloss} for recent reviews). 
There seems to be a general consensus, though, that neither of these solves the problem (at least in the form stated above); they rather reconfirm it. Indeed,  granted that measurements yield classical data,  since classical physics by definition  does not have quantum-mechanical interference terms, their disappearance in appropriate limiting situations (like the ones described by decoherence and/or superselection theory) 
 is just a necessary condition that (in part) \emph{defines} measurement (which after all is supposed  to produce some classical state as its outcome).\footnote{In this sense even von Neumann's book \cite{vN32} is misleading, since he suggested that the act of observation may be identified with a cut in the chain now named after him. What is right about this idea is that observation is linked to a voluntary change of information, but it would have been preferable to point out  that such a change, in so far as it defines measurement, should be a \emph{loss of quantum information}.} Consequently, in our opinion the \mmp\ is \emph{posed}, rather than  \emph{solved} by proving that such interference terms vanish under particular  (limiting) conditions.\footnote{Furthermore, despite its outspoken ambition to derive classical physics from quantum theory \cite{Joos,Schloss,Zurek1}, decoherence hardly (if at all) invokes limits like Planck's constant going to zero, which are needed, for one thing, to derive the correct classical equations of motion
 (cf.\ \cite{handbook,KlaasSchloss}). }

In contrast, the  \emph{dynamical collapse models} of Pearle, Ghirardi--Rimini--Weber,  and others (cf.\ \cite{BG} for a comprehensive survey) \emph{do} solve the measurement problem. But they do so at a price: the Schr\"{o}dinger equation is modified by adding a novel and universal stochastic process that even makes the equation nonlinear, and which, like the solution of Heisenberg and von Neumann, is  ad hoc except for its goal of causing collapse. 

The approach to the measurement problem we are going to propose below uses key ideas from both dynamical collapse models and decoherence (and could not have been conceived without the inspiration from these earlier approaches), but in such a way
that we avoid some of their drawbacks (though only in the context of a simple example, so far):
\begin{enumerate}
\item Dynamical collapse is obtained without modifying quantum theory.
\item While decoherence preserves all peaks (i.e., potential measurement outcomes) in the density matrix, and hence subsequently needs e.g.\ some kind of a Many Worlds Interpretation \cite{Han, Schloss,Wallace}, our mechanism, if correct, leads to just one outcome.
\end{enumerate}
Our approach starts with a technical reformulation of the  measurement problem, which relies on a specific mathematical formalism for dealing with classical states, including their role as potential limits of quantum states. For completeness' sake, we explain this first. 
\subsection{Intermezzo: classical states}\label{classicalstates}
In order to describe classical states as limits of quantum states,  we need to  describe all states algebraically.\footnote{An introduction for philosophers  to the material in this section may be found in \cite[Ch.\ 4\&5]{handbook}.} Although this formalism implicitly uses the language of C*-algebras, very little of that theory will be needed here \cite{Haag,book}. The main point is to have a unified formalism for classical and quantum physics, from which the usual treatment in which classical physics is described using phase spaces having the structure of (differentiable) manifolds, whilst quantum physics is based on \Hs s and operators, is a far cry. 

Nonetheless, the necessary unification of these diverse instances of mathematical formalism may be achieved by focusing on the \emph{interplay} between observables (or propositions) and states (as opposed to the \emph{reification} of the notion of a state inherent in the usual approach, which is particularly detrimental to the  \mmp). Indeed, classical and quantum physics turn out to share the following mathematical structure:
\begin{itemize}
\item The observables comprise the (self-adjoint part of) a C*-algebra $A$.\footnote{A C*-algebra is a complex algebra $A$ that is complete in a norm $\|\cdot\|$ satisfying $\| ab\|\,\leq\, \| a\|\,\|
b\|$ for all $a,b\in A$, and has an involution $a\raw a^*$ such that $\| a^*a\|=\| a\|^2$.}
\item  The states are positive linear functionals  $\om:A\raw\C$  of norm one on $A$.\footnote{Positivity of  $\om$ means that $\om(a^*a)\geq 0$ for all $a\in A$.
 If $A$ has a unit $1$, then a state may equivalently be defined as a  positive linear functional $\om:A\raw\C$ that satisfies $\om(1)=1$.} 
 \item A \emph{pure} state $\om$ has no nontrivial convex decomposition, i.e., if $\om=p\om_1+(1-p)\om_2$ for some $p\in (0,1)$ and certain states $\om_1$ and $\om_2$, then $\om_1=\om_2=\om$.
\end{itemize}
In the main example of interest for this paper, which is a particle moving on the real line, this abstract talk has a  simple implementation (which suffices  for what follows). Its main feature is that, as Heisenberg had it, the difference between the classical and the quantum setting  lies in the non-commutativity of the observables in the latter (see also \cite{CBH}).
\begin{description}
\item[Classical:] The C*-algebra of observables is $A_0=C_0(\R^2)$, that is, the continuous functions on the phase space $\R^2$ that vanish at infinity.
The algebraic operations are defined pointwise (e.g., $(fg)(z)=f(z)g(z)$), involution is (pointwise) complex conjugation, and the norm is the supremum-norm. Hence $A_0$ is commutative. A state  $\mu:A_0\raw\C$ is essentially the same thing as a probability measure $\hat{\mu}$ on phase space through
\beq\mu(f)=\int_{\R^2} d\hat{\mu}\, f.
\eeq
Such a probability  \emph{measure} may or may not be given by a probability  \emph{density} with respect to the Liouville measure $dpdq/2\pi$, i.e., a positive $L^1$-function $\chi$ on $\R^2$ s.t.
\begin{equation}
\int_{\R^{2n}} \frac{dp
dq}{2\pi}\, \chi(p,q) f(p,q)=
\int_{\R^2} d\hat{\mu}\, f. \label{maymaynot}
\end{equation}
This also implies that $\chi$  integrates to unity (with respect to the Liouville measure).
Pure states on $C_0(\R^2)$ are probability measures of the Dirac form $\dl_z$, $z\in\R^2$, i.e., $\dl_z(f)=f(z)$ for $f\in C_0(\R^2)$,  and hence bijectively  correspond to
 points of $\R^2$. Such states are not given by probability densities, however, as $\dl$-functions are not in $L^1$. 
\item[Quantum:] The C*-algebra of observables is $A=K(L^2(\R))$ i.e., the algebra of compact operators on the \Hs\ $L^2(\R)$ of square-integrable wave-functions.\footnote{A self-adjoint operator $a$ on a \Hs\ is compact just in case it has a spectral decomposition $a=\sum_i \lm_i |e_i\ra \la e_i|$ (where $|e_i\ra \la e_i|$ is the orthogonal projection onto the ray $\C e_i$), where  the eigenvectors $(e_i)$ form an orthonormal basis of $H$, each nonzero eigenvalue $\lm_i$ has finite multiplicity, and  if the eigenvalues are listed in decreasing order of their absolute value (i.e.,
$\|a\|=|\lm_1|\geq |\lm_2| \geq \cdots$), then $\lim_{\al\raw\infty} |\lm_{\al}|=0$. } 
Such operators may be added and multiplied in the obvious way; involution is hermitian conjugation, and the norm is the usual operator norm.
States
$\rh:K(L^2(\R))\raw\C$  bijectively correspond to density matrices $\hat{\rh}$ on $L^2(\R)$ through  $\rh(a)= \Tr (\hat{\rh} a)$.  
A unit vector $\Ps$ defines a  density matrix $|\Ps\rangle\langle\Ps|$, so that
pure states $\psi$ on $K(L^2(\R))$ bijectively correspond to unit vectors $\Ps$ up to a phase (or to rays $\C\cdot\Ps$)  through
 \beq
 \ps(a)=\la\Ps|a|\Ps\ra\equiv \la\Ps, a\Ps\ra. \label{ps}
 \eeq
\end{description}
In both cases, an idealized description has been used in order to avoid unnecessary mathematical complications: 
first, inherently to the C*-algebraic formalism, one works with \emph{bounded} operators (in quantum theory) and functions (classically).\footnote{Although unbounded operators play a major practical role in physics (think of position and momentum), they may be awkward to deal with due to domain issues and in any case they can theoretically be avoided without any loss of generality. Indeed, unbounded self-adjoint operators $a$ bijectively correspond to bounded operators (or constructions involving those) in at least four different ways \cite{RS1}. First, one may pass to the unitary Cayley transform $(a-i)(a+i)\inv$. Second, one may construct the associated one-parameter unitary group $t\mapsto \exp(ita)$ by Stone's Theorem. Third, one may work with the bounded spectral projections, from which the operator may be reconstructed by the spectral theorem. Fourth, one could take the resolvents $(a-\lm)\inv$, $\lm\notin\R$ (whose typical integral kernels are Green's functions).} Second, the restriction to functions vanishing at infinity and compact operators in the classical and the quantum case, respectively, is also purely a matter of mathematical convenience.\footnote{Classically, one could work with the (essentially) bounded integrable functions
$L^{\infty}(\R^2)$, whilst in quantum theory one could take the algebra $B(L^2(\R))$ of all bounded operators on $L^2(\R)$ (as opposed to merely the compact ones). These broader classes can be constructed from the $C_0$-functions or compact operators by taking pointwise and weak (or strong)  limits, respectively.  Enlarging the class of observables like that also brings in new continuity conditions on the states (namely $\sg$-additivity and $\sg$-weak continuity, respectively). Imposing these leads to the same states as discussed above; without them, one obtains more.}

The following notion of convergence of quantum states to classical ones is standard (cf.\ \cite{CP,book,PU,Robert} and many other sources),\footnote{Often Weyl quantization $Q^W_{\hbar}$ is used instead of Berezin quantization $Q_{\hbar}$, as in \cite{BB},
but for Schwartz functions $f$ on phase space these have the same asymptotic properties as 
$\hbar\raw 0$ \cite{book}. The advantage of Berezin quantization is that it is well defined also for continuous functions vanishing at infinity, in that for any unit vector $\Ps\in L^2(\R)$ the map $f\mapsto \la \Ps| Q_{\hbar}(f)|\Ps\ra$ defines a probability measure on phase space. In contrast, the Wigner function 
defined by $f\mapsto \la \Ps| Q^W_{\hbar}(f)|\Ps\ra$ may fail to be positive, as is well known.}  and has been used especially in quantum chaology \cite{NonVoros}. We first  recall the \emph{coherent states}, labeled by $z=(p,q)\in\R^2$, 
\beq
\Phi^{(p,q)}_{\hbar}(x)=(\pi\hbar)^{-1/4}e^{-
ipq/2\hbar}e^{ipx/\hbar}e^{-(x-q)^2/2\hbar},\label{pqcohst} 
\eeq
with associated \emph{Berezin quantization} map $f\mapsto Q_{\hbar}(f)$, $f\in C_0(\R^2)$, $ Q_{\hbar}(f)\in K(L^2(\R))$,
\beq
 Q_{\hbar}(f)=\int_{\R^{2n}} \frac{dp
dq}{2\pi\hbar}\, f(p,q) | \Phi^{(p,q)}_{\hbar}\rangle\langle\Phi^{(p,q)}_{\hbar}|.
\eeq
Now let $(\rh_{\hbar})$ be a family of quantum states, indexed by $\hbar$ (say $\hbar\in (0,1]$), 
with associated density matrices $(\hat{\rh}_{\hbar})$, and let $\rh_0$ be a state on $C_0(\R^2)$, with 
associated
probability measure $\hat{\rh}_0$ on $\R^2$. The quantum states $(\rh_{\hbar})$ \emph{converge} to the classical state $\rh_0$  if for all  $f\in  C_0(\R^2)$,
\beq
\lim_{\hbar\raw 0} \rh_{\hbar}(Q_{\hbar}(f))=\rh_0(f). \label{climit}
\eeq 
In that case we write  $\lim_{\hbar\raw 0} \rh_{\hbar}=\rh_0$.
The condition \er{climit} more explicitly reads 
\begin{equation}
\lim_{\hbar\raw 0}  \Tr (\hat{\rh}_{\hbar}  Q_{\hbar}(f))=\int_{\R^2} d\hat{\rh}_0\, f, \: \mbox{ for all }f\in C_0(\R^2).
\end{equation}
If $\rh_{\hbar}=|\Ps_{\hbar}\rangle\langle\Ps_{\hbar}|$, then obviously
\begin{equation}
\langle\Ps_{\hbar}| Q_{\hbar}(f)|\Ps_{\hbar}\rangle=\int_{\R^{2n}} \frac{dp
dq}{2\pi\hbar}\, \chi_{\Ps_{\hbar}}(p,q)
 f(p,q) ,
\end{equation}
where the probability density $\chi_{\Ps_{\hbar}}$, called the \emph{Husumi function} of $\Ps_{\hbar}$, is given by 
\begin{equation}
 \chi_{\Ps_{\hbar}}(p,q)=|\langle \Phi^{(p,q)}_{\hbar}|\Ps_{\hbar}\rangle|^2,
\end{equation}
in which  the inner product is  taken in $L^2(\R)$. 
Consequently, if the limit in \er{climit} exists for specific $\rh_{\hbar}=|\Ps_{\hbar}\rangle\langle\Ps_{\hbar}|$, then the limit measure $\hat{\rh}_0$ is the weak (or pointwise) limit of the probability measures $\mu_{\Ps_{\hbar}}$
defined by the probability densities $\chi_{\Ps_{\hbar}}$ according to \er{maymaynot}.

\begin{figure}[H]
\begin{center}
\includegraphics[width=0.49\textwidth]{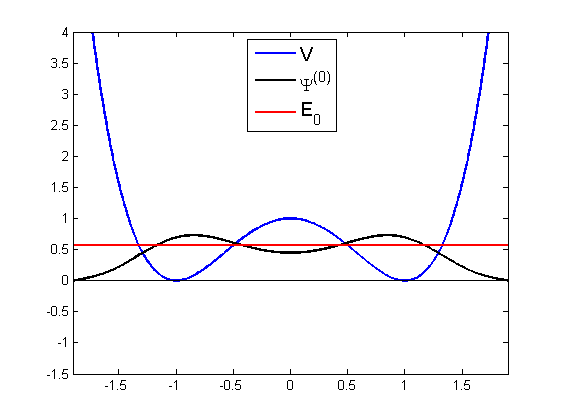}
\includegraphics[width=0.49\textwidth]{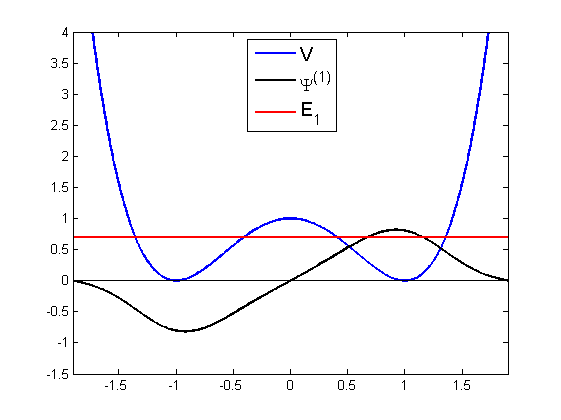}
\caption{Double-well potential with ground state $\Ps_{\hbar=0.5}^{(0)}$ and first excited state $\Ps_{\hbar=0.5}^{(1)}$.}
\label{QMDubbelePut}
\end{center}
\end{figure}

Let us illustrate this formalism for the  ground state  of the  one-dimensional harmonic oscillator. Taking $V(x)=\half \om^2x^2$   
in  the usual quantum Hamiltonian (with mass $m=1/2$),
\begin{equation}
H_{\hbar}=-\hbar^2\frac{d^2}{dx^2} +V(x), \label{TheHam}
\end{equation}
it is well  known that the ground state is unique  and that its wave-function  
 \beq
 \Ps_{\hbar}^{(0)}(x)=\left(\frac{\om}{2\pi\hbar}\right)^{1/4}e^{-\om x^2/4\hbar}
 \eeq
  is a Gaussian, peaked above $x=0$. As $\hbar\raw 0$, this ground state converges  to the ground state $\rh_0^{(0)}=(0,0)\in\R^2$ (i.e., $(p=0,q=0)$) of the corresponding classical system. 
Slightly less familiar, the same is true for the
 \emph{an}harmonic oscillator (with small  $\lm>0$), i.e., 
\begin{equation}
V(x)=\half \om^2x^2+\quar \lm x^4,\label{Vq2}
\end{equation}
the peak, of course,  now being only approximately Gaussian.
But it is a deep and counterintuitive feature of quantum theory that even the symmetric double-well potential
\begin{equation}
V(x)=-\half \om^2x^2+\quar \lm x^4+\quar\om^4/\lm=\quar \lm (x^2-a^2)^2 , \label{Vq3}
\end{equation}
where $a=\om/\sqrt{\lm}>0$ (assuming $\om>0$ as well as $\lm>0$), has a unique quantum-mechanical  ground state    \cite{HiSi,RS4},
despite the fact that the corresponding classical system has two degenerate ground states, given by the phase space  points $\rh_0^{\pm}\in\R^2$ defined by 
\beq
\rh_0^{\pm}=(p=0,q=\pm a). \label{clgrstates}
\eeq 
This ground state wave-function $\Ps_{\hbar}^{(0)}$  is again real and positive definite, but this time it has \emph{two} peaks, above $x=\pm a$, with exponential decay $|\Ps_{\hbar}^{(0)}(x)|\sim \exp( -1/\hbar)$ in the classically forbidden region  \cite{HiSi,RS4}. 
As a quantum-mechanical shadow of the classical degeneracy, energy eigenfunctions (and the associated eigenvalues) come in pairs. In what follows, we will be especially interested in the first excited state $\Ps_{\hbar}^{(1)}$, which like $\Ps^{(0)}_{\hbar}$ is real, but has one peak \emph{above}  $x=a$ and another peak \emph{below} $x=-a$.
See Figure \ref{QMDubbelePut}. 

\begin{figure}[H]
\begin{center}
\includegraphics[width=0.49\textwidth]{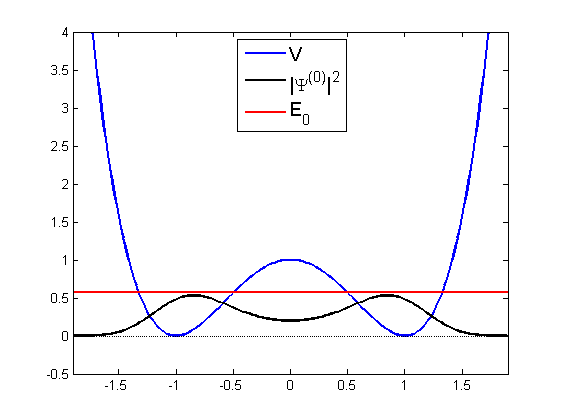}
\includegraphics[width=0.49\textwidth]{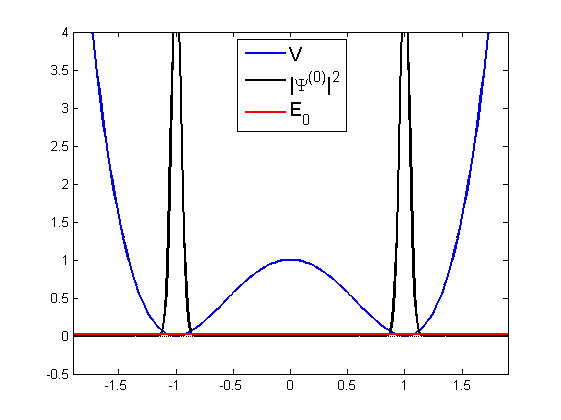}
\caption{Probability densities for  $\Ps_{\hbar=0.5}^{(0)}$ (left)  and $\Ps_{\hbar=0.01}^{(0)}$ (right).}
\label{PD}
\end{center}
\end{figure}

As $\hbar\raw 0$,  the eigenvalue splitting $E_1-E_0$ vanishes exponentially in $-1/\hbar$ like
\beq
\Dl\equiv E_1-E_0\sim(\hbar\om/\sqrt{\half e\pi})\cdot e^{-d_V/\hbar} \:\: (\hbar\raw 0), \label{ED}
\eeq
where the typical {\sc wkb}-factor is given by
\beq
d_V=\int_{-a}^a dx\, \sqrt{V(x)}; \label{Cwkb}
\eeq
see \cite{Garg,LL} (heuristic), or \cite{Helffer,HiSi,Simon4} (rigorous)  for details. Also,  the probability density of each of the wave-functions $\Ps^{(0)}_{\hbar}$ or $\Ps_{\hbar}^{(1)}$ contains approximate  $\delta$-function peaks above \emph{both} classical minima $\pm a$. See  Figure \ref{PD}, displayed just for $\Ps^{(0)}_{\hbar}$, the other being analogous.

\begin{figure}[H]
\begin{center}
\includegraphics[width=0.49\textwidth]{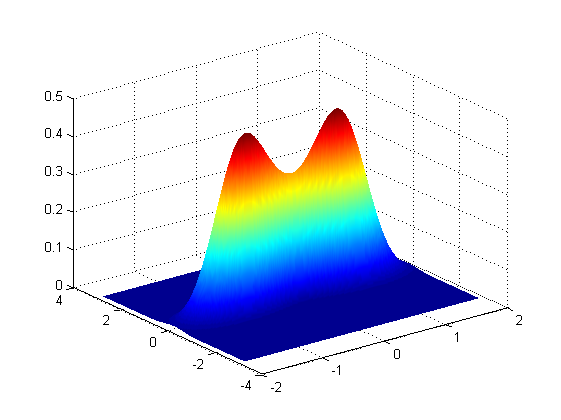}
\includegraphics[width=0.49\textwidth]{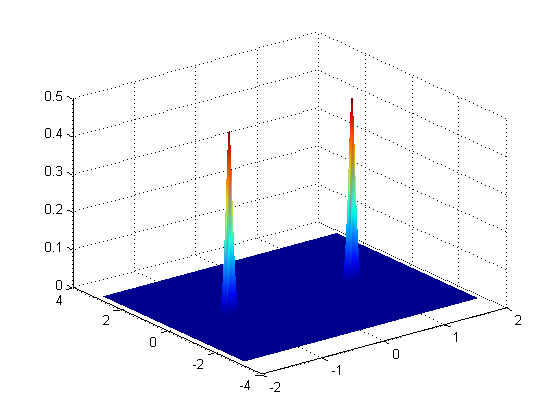}
\caption{Husumi functions for $\Ps_{\hbar=0.5}^{(0)}$ (left) and  $\Ps_{\hbar=0.01}^{(0)}$ (right).}
\label{HF}
\end{center}
\end{figure}

We can make the correspondence between the \emph{nondegenerate} pair $(\Ps^{(0)}_{\hbar},\,\Ps_{\hbar}^{(1)})$  of low-lying quantum-mechanical  wave-functions and the pair $(\rh_0^+,\rh_0^-)$ of \emph{degenerate} classical ground states  more transparent  by invoking the above notion of a classical limit.  Indeed, in terms of the corresponding algebraic states $\psi^{(0)}_{\hbar}$ and  $\psi_{\hbar}^{(1)}$, cf.\ \er{ps},  one has
\begin{eqnarray}
{\lim_{\hbar\raw 0} \psi_{\hbar}^{(0)}}&=&\lim_{\hbar\raw 0} \ps^{(1)}_{\hbar}=\rh_0^{(0)}, \label{1.5} \\
\rh_0^{(0)}&\equiv& \half(\rh_0^+ + \rh_0^-), \label{1.6}
\end{eqnarray}
where $\rh_0^{\pm}$ are the pure classical ground states \er{clgrstates} of the double-well Hamiltonian.\footnote{In \er{1.6}  we regard classical states as probability measures on phase space; hence the addition on the right-hand side is a convex sum of measures, which has nothing to do with addition in the particular phase space $\R^2$ (whose linear structure is accidental and irrelevant). } 
To see this, one may either consider numerically computed Husumi functions, as  shown in Figure \ref{HF} (just for $\Ps^{(0)}_{\hbar}$, as before), 
or proceed analytically, combining the relevant estimates in \cite{Harrell} or in \cite{Simon4} with the computations in \S II.2.3 of \cite{book}.
Either way, it is clear that
 the \emph{pure} (algebraic) quantum ground state $\psi_{\hbar}^{(0)}$
 converges to the \emph{mixed} classical state \er{1.6}.
 
In contrast with $\Ps_{\hbar}^{(0)}$ and $\Ps_{\hbar}^{(1)}$, the localized (but now time-dependent) wave-functions
\begin{equation}
\Ps_{\hbar}^{\pm}= \frac{\Ps_{\hbar}^{(0)}\pm\Ps_{\hbar}^{(1)}}{\sqrt{2}}, \label{plusmin}
\end{equation}
which of course define pure (algebraic) states as well, converge to \emph{pure} classical states, i.e., 
\beq
\lim_{\hbar\raw 0} \ps^{\pm}_{\hbar}=\rh_0^{\pm}.\label{1.7}
\eeq
On the one hand this is not surprising, because $\Ps_{\hbar}^{\pm}$ has a single peak above $\pm a$, but on the other hand it is, since neither $\Ps_{\hbar}^+$ nor $\Ps_{\hbar}^-$ is an energy eigenstate (whereas their limits $\rh_0^+$ and $\rh_0^-$ \emph{are} energy eigenstates, in the classical sense of being fixed points for the Hamiltonian flow). The explanation is that the energy difference \er{ED} vanishes exponentially as $\hbar\raw 0$, so that  in the classical limit $\Ps_{\hbar}^+$ and $\Ps_{\hbar}^-$ approximately do become energy eigenstates. In similar vein, because of \er{ED} the tunneling time  $\ta=2\pi\hbar/\Dl$ of the oscillation between $\Ps_{\hbar}^+$ and $\Ps_{\hbar}^-$  becomes exponentially large in $1/\hbar$ as $\hbar\raw 0$.

Finally, in the above examples (and many others) time evolution of states  is defined both classically (by the Liouville equation for measures) and quantum mechanically (by the von Neumann equation for density matrices), and provided that $\lim_{\hbar\raw 0}\rh_{\hbar}=\rh_0$ as in \er{climit},
time-evolution commutes with taking the classical limit: that is, 
for each fixed time $t\in\R$, one has Egorov's Theorem in the form \cite[Thm.\ II.2.7.2]{book}, \cite{Robert}
\beq
\lim_{\hbar\raw 0} (\rh_{\hbar}(t))=\rh_0(t). \label{egorov}
\eeq
\newpage

\newpage
 \subsection{Reformulation of the \mmp}\label{sec2.3}
 We return to the \mmp.
If measurement is merely seen as the establishment of certain correlations between two quantum systems, then the problem does not arise, since \emph{a priori} nothing is wrong with the existence of superpositions of such correlated quantum states. What \emph{is} wrong is that at first sight such superpositions seem to survive the classical limit, as shown above by the ground state of the double-well potential. More generally,  the \mmp\ arises 
whenever in the classical limit  a \emph{pure} quantum state converges to a \emph{mixed} classical state, since in that case quantum theory fails to predict a single measurement outcome. Rather, it suggests there are many outcomes, \emph{not} just because the wave-function has several peaks \emph{per se}, but because \emph{in addition} in the classical limit each of these peaks converges to a different classical state.

Consequently, the \mmp\ is by no means  solved by proving that such interference terms vanish under certain (limiting) conditions.
Instead, the real problem is to show that under realistic measurement conditions pure quantum states actually have \emph{pure} classical limits. Indeed, Schr\"{o}dinger's Cat is exactly of this nature \cite[\S 6.6]{handbook}:
\begin{itemize}
\item If one were to study the cat as a quantum system, nothing would be  wrong with the famous superposition it resides in. However, such a study is practically impossible.\footnote{This appeal to ``practice'' does not mean that we are resigned 
to {\sc fapp} (i.e., ``for all practical purposes'') solutions to the \mmp. As in \cite{KlaasObs}, we remain convinced
that the classical description of a measurement apparatus is a purely epistemic move, relative to which 
outcomes are \emph{defined}. So even if it were possible to study a cat  as a quantum system, there would be no \mmp, since in that case there would be innumerable superpositions but not a single (undesirable) mixture of classical states. 
}
\item The ``paradox'' arises only  if one uses macroscopic variables  in order to give a classical description of the cat, so that notions like (being) ``alive'' or ``dead'' make sense. In that case, the naive classical state of the cat is of the kind $\rh_0^{(0)}=\half(\rh_0^+ + \rh_0^-)$, cf.\ \er{1.6}, where (say) $\rh_0^+$ stands for being alive and
 $\rh_0^-$ is the (classical) state of death. A classical state like $\rh_0^{(0)}$ is indeed intolerable, but since our flea destabilizes it, fortunately enough it cannot arise in practice (in theory, such a state could be created in a totally isolated system, in which case its paradoxical features disappear).
\end{itemize}

Having   endorsed the Bohrian (or rather: the experimentally relevant) view of what a measurement is, we emphatically \emph{reject} the (typically) accompanying claims that the measurement process itself cannot be analyzed or described in principle, and that its outcome is irreducibly random (except for special initial states).\footnote{We share this rejection with the Bohmians \cite{Cushing}.
The folk wisdom (shared by the Founding Fathers) that the Copenhagen Interpretation has no measurement problem relies on these secondary Copenhagenian claims, which indeed sweep the problem under the rug. Incidentally, these claims seem much more popular than Bohr's doctrine of classical concepts, which is generally not well understood, and/or mistaken for the idea that the goal of physics is to explain experiments, or that reality does not exist, et cetera. }
For if measurement \emph{by definition} produces some classical state from a quantum state, and
quantum (field) theory is agreed to be fundamental and hence classical physics is some limit of it \cite{book,handbook},  then it would seem almost perverse  not to describe the pertinent limiting procedure explicitly. 

Take our example of the classical limit of the double-well potential, cf.\  \er{plusmin} etc. Its ground state $\Ps_{\hbar}^{(0)}=(\Ps_{\hbar}^{+}+\Ps_{\hbar}^{-})/\sqrt{2}$ is supposed to model the combined post-measurement state $\Psi$ of  some apparatus $A$ coupled to a  microscopic system $S$, where the latter was initially prepared in a superposition $(\Phi_+ +\Phi_-)/\sqrt{2}$,
upon which the measurement interaction brought $S+A$ into the state $(\Phi_+\ot\Om_++\Phi_-\ot\Om_-)\sqrt{2}$, cf.\ \S\ref{histo}, so that
$\Ps_{\hbar}^{\pm}\simeq\Phi_{\pm}\ot\Om_{\pm}$.

\noindent We then
 interpret the double limit $\hbar\raw 0$, $t\raw\infty$  (in  an appropriate order discussed below) as the ``unfolding'' of the  measurement, in that  the apparatus $A$ is described increasingly classically whilst $S$ disappears from the description.\footnote{The analogy with the thermodynamic limit  will be discussed in  \S\ref{SBPT}.
As to the limit $\hbar\raw 0$, we repeat \cite[pp.\ 471--472] {handbook} that although  $\hbar$ is a dimensionful  \emph{constant}, in practice one studies the (semi)classical regime of a given quantum theory by forming a dimensionless combination of $\hbar$ and other parameters; this combination then re-enters the theory as if it were a dimensionless version of  $\hbar$ that can indeed be varied. The oldest example is Planck's radiation formula, with the associated limit $\hbar\nu/kT\raw 0$, and 
another example is the Schr\"{o}dinger operator \er{TheHam}, with mass reinserted, where  one may pass to a dimensionless parameter $\hbar/\lm\sqrt{2m\epsilon}$,
where $\lm$ and  $\epsilon$ are typical length and energy scales, respectively.}
What happens in this process?
 \begin{itemize}
\item  According to the Copenhagen Interpretation, by some inexplicable mystery, at some stage of the classical description  the wave-function suddenly collapses.
\item According to our analysis, the collapse is not inexplicable at all: it  is caused by a perturbation, and  in principle it can be exactly described and followed in time.
\end{itemize}
Here it  is important to note that the values $\hbar=0$ and $t=\infty$ are never actually reached: we are talking about \emph{limits}! In particular, the instability of the ground state described in the next section already arises for \emph{very small} (as opposed to zero or `infinitesimal') effective values of $\hbar$. And this is how it should be: truly classical states (like strictly infinite systems) do not exist in nature, but you should be able to make the difference between
 the quantum-mechanical approximation to such a state and the actual limit state
 as small as you like,  for sufficiently small $\hbar$ 
 and  large $t$.
Indeed, the whole point  is that 
the usual (superposition) state $\Ps^{(0)}_{\hbar}$ of Schr\"{o}dinger's Cat does \emph{not} have this feature: the classical states that (almost) occur in nature are $\rh_0^+$ (alive) and  $\rh_0^-$ (dead), and for any $\hbar>0$, the state  $\ps^{(0)}_{\hbar}$ defined by the wave-function $\Ps^{(0)}_{\hbar} $ dramatically fails to approximate either of
 these,\footnote{Paraphrasing Bell \cite{BellCH}: the difference between $\rh_0^{\pm}$ and $\psi_{\hbar}^{(0)}$ can be made `as big as you do \emph{not} like.'. } although it perfectly well approximates the unphysical mixture $\half(\rh_0^+ + \rh_0^-)$. In other words, returning to the original mechanical meaning of the double-well system, quantum mechanics is apparently unable to predict that a classical ball 
lies at the bottom of either  the right or  the left well. 
 Fortunately, and this is the key to our analysis of the \mmp, this inability is only apparent: 
 depending on the sign and localization of the perturbation $\dl V$ of the double well (cf.\ the next section), the collapsed states $\ps^{(\dl)}_{\hbar}$ induced by the ``flea on the cat'' \emph{do} approximate either $\rh_0^+$ or  $\rh_0^-$ as $\hbar\raw 0$.
 
 However, this insight concerning perturbed ground states and their associated localized wave-functions is only the first, \emph{static}  part
of the solution of the \mmp. The \emph{dynamical} part of the solution would be to find an appropriate time-dependent way for the flea to jump 
onto Schr\"{o}dinger's Cat (in its superposition state), and either kill it, or let it live. That is, one needs to find a
suitable  perturbed (but nonetheless  unitary!) quantum  time-evolution operator $U_{\hbar}^{(\dl)}(t)$ such that  the (algebraic) state
$\psi^{(\delta)}_{\hbar}(t)$ defined by the wave-function $U_{\hbar}^{(\dl)}(t)\Ps^{(0)}_{\hbar}$ converges to either $\rh_0^+$ or  $\rh_0^-$ as $t\raw\infty$ and $\hbar\raw 0$. Moreover, a completely  satisfactory solution of the measurement problem (or at least of its  Schr\"{o}dinger Cat instance) would 
have the additional property that measurement results that are already pre-classical, which in this case means that they are either $\ps^+_{\hbar}$ or $\ps^-_{\hbar}$, be stable under perturbations.
 This leads to the following conditions:
 
\begin{eqnarray}
\lim_{\hbar\raw 0,t\raw\infty} \ps^{(0)}_{\hbar}(t)&=& \mbox{ either } \rh_0^+ \mbox{ or }  \rh_0^-; \label{solcat}\\
{\lim_{\hbar\raw 0,t\raw\infty} \ps^+_{\hbar}(t)}&=& \rh_0^+;\label{solcat2} \\
{\lim_{\hbar\raw 0,t\raw\infty} \ps^-_{\hbar}(t)}&=& \rh_0^-. \label{solcat3}
\end{eqnarray}
As in \er{1.5} - \er{1.7},  these conditions  do not contradict each other, since in passing from unit vectors $\Ps$ to algebraic states $\psi$ (which are quadratic in $\Ps$), \emph{linearity is lost}. Indeed, 
\er{1.5} - \er{1.6} would be impossible for unit vectors (noting that $\Ps_{\hbar}^{(0)}$ and $\Ps_{\hbar}^{(1)}$ are orthogonal), but they are perfectly alright for algebraic  states.\footnote{Families of unit vectors like $\Ps^{(i)}_{\hbar}$, where $i=0,1,+,-$, typically do not have a limit as unit vectors (or even as density matrices, including one-dimensional projections).} This marks a decisive difference with 
standard approaches to the measurement problem  \cite{BG, Bub,BLM},  which 
 are victim to ``insolubility'' theorems of the kind proved by Fine and others \cite{Brown,BLM,Fine}. Such theorems assume that the post-measurement state (if pure) is a unit vector in Hilbert space (or a density matrix otherwise), and totally rely on the linearity of the Schr\"{o}dinger equation. In contrast, we  take  the post-measurement state to be classical \emph{by definition}.
 
 In the above setting, determining the correct way to take the double limit $\hbar\raw 0,t\raw\infty$ is a highly nontrivial problem. In the theory of 
 semiclassical asymptotics \cite{Robert} (with quantum chaology as an important subfield \cite{BDB,Faure,Schubert,Zas}), the goal of this limit is to find the long-time behaviour of some quantum system by first describing the underlying classical system (especially if it is chaotic), and subsequently using suitable classical expressions to approximate the corresponding quantum formulae. For example, suppose one wants to find the time evolution of a wave-packet that initially is strongly (micro)localized, i.e., is a coherent state for some small (effective) value of $\hbar$.  For fixed time $t$, one has Egorov's Theorem \er{egorov}, which, supplemented by exponentially small error terms, shows that for any finite $t$ the limit $\hbar\raw 0$ delivers the above goal.  But for large times there is a competition between the limit $\hbar\raw0$ making the state more localized (and hence more classical), and
 the limit  $t\raw\infty$ making it less so (and hence more wave-like or quantum-mechanical).  Intuitively, spreading is enhanced if the classical dynamics is chaotic, and suppressed if it is integrable. In the chaotic case, it turns out that (micro)localization defeats the spread in time as long as $t\leq C \ln(1/\hbar)$, with $C$ of order one, so that one may take the double limit in the order $\hbar\raw 0, t\raw  C \ln(1/\hbar)$ \cite{BGP,BR,CR}. 
 If the system is integrable, on the other hand, one expects to push this to much larger times $t\sim \hbar^{-k}$, for some $k\in\N$.
 
Our situation is more complicated than that. First,  in $d=1$,  time-dependent perturbations of the ``flea'' type render the double-well potential no longer integrable, without the perturbed dynamics becoming really chaotic either. Second, the initial state is the ground state of the (unperturbed) double well, which is not even localized to begin with.\footnote{As explained above, the nonlinearity inherent in the limit $\hbar\raw 0$ makes it impossible to find the limit of this ground state $\Ps^{(0)}_{\hbar}$  by just adding the results for two localized wave-functions like $\Ps^+_{\hbar}$ and $\Ps^-_{\hbar}$.}
Third, we will actually invoke another limit, namely the adiabatic one. As we will explain in the Discussion, combining these features poses a new problem in the practically unexplored territory of quantum metastability, whose solution will not only involve new mathematical results in 
semiclassical asymptotics, but also calls for genuinely new physical understanding. 
 For now, our goal is just to explain our program and provide a ``proof of concept'' that it might work. 
Thus at the present stage we merely present some numerical results, showing that for fixed small $\hbar$, localization takes place for sufficiently large $t$.
\section{A collapse process within \qm} \label{FE}
\subsection{The ``flea'' perturbation of the double-well potential}\label{sec3.1}
Regarding the doubly-peaked ground state $\Ps_{\hbar}^{(0)}$ of the  symmetric double well as the quantum-mechanical counterpart of a hung parliament, the analogue of a small party that decides which coalition is formed is a tiny \emph{asymmetric} perturbation $\dl V$ of the potential. Indeed, the following spectacular phenomenon in the theory of Schr\"{o}dinger operators was  discovered  in 1981 by Jona-Lasinio, Martinelli and Scoppola \cite{JLMS1,JLMS2}, using stochastic techniques.  Using more conventional methods, it was subsequently reconfirmed and analyzed further by  mathematical physicists  \cite{CDS2,GGJL,Helffer,HS,Simon4}.\footnote{The ``Flea on the Elephant'' terminology used in  \cite{Simon4} for the phenomenon in question evidently motivated the title of the present paper, which has identified the  proper host animal at last!}  In view of this extensive mathematical literature, we hardly see a need for yet another  rigorous treatment, but rather take it as our goal to explain the main idea to physicists and philosophers. This section just gives the key results; a more detailed (an novel) treatment using the well-known {\sc wkb} approximation from the textbooks may be found in the appendix. 

\begin{figure}[H]
\begin{center}
\includegraphics[width=0.98\textwidth]{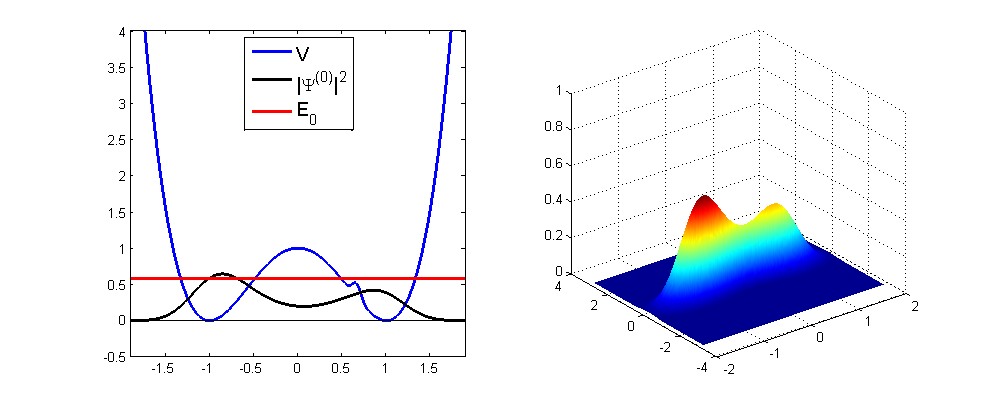}
\caption{Flea perturbation of ground state $\Ps_{\hbar=0.5}^{(\dl)}$ with corresponding Husumi function. For such relative large values of $\hbar$, little (but some) localization takes place.}
\label{Fleap1}
\end{center}
\end{figure}

\begin{figure}[H]
\begin{center}
\includegraphics[width=0.98\textwidth]{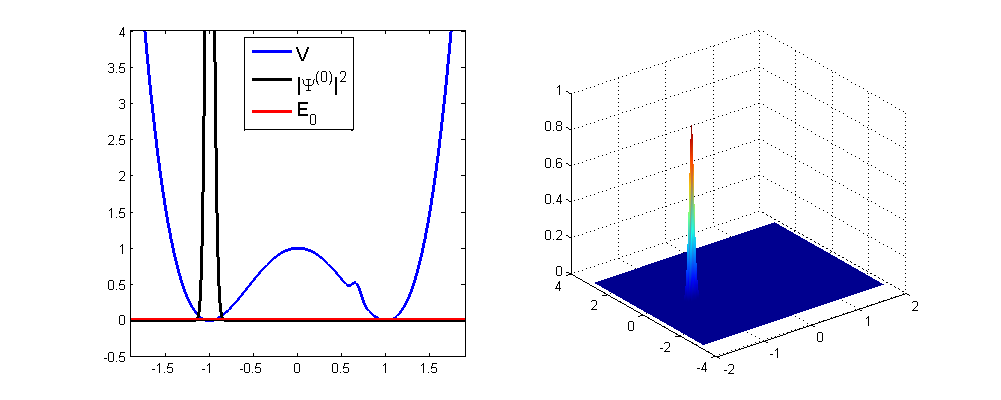}
\caption{Same at $\hbar=0.01$. For such small values of $\hbar$, localization is  almost total.}
\label{Fleap2}
\end{center}
\end{figure}

\begin{figure}[H]
\begin{center}
\includegraphics[width=0.98\textwidth]{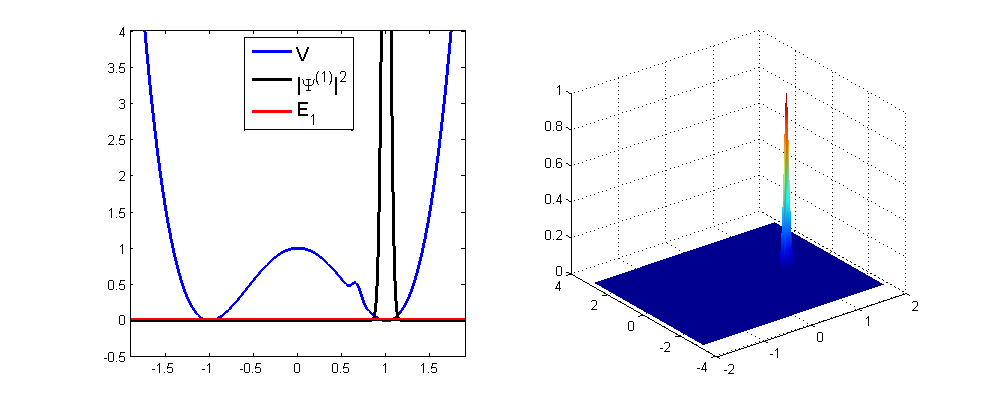}
\caption{First excited state for $\hbar=0.01$. Note the opposite localization area.}
\label{Fleap3}
\end{center}
\end{figure}

Replace $V$ in \er{TheHam} by $V+\dl V$, where $\dl V$ (i.e., the ``flea'')  is assumed to:
\begin{enumerate}
\item be real-valued with fixed sign, and $\cci$ (hence bounded) with connected support  not including the minima $x=a$ or $x=-a$;\footnote{
Some of the details in this section depend on the latter assumption, but our overall scenario in Section \ref{Dynamical} does not. 
For example, if the value and/or the curvature of one of the minima is decreased, then the ground state wave-function will localize above that minimum, as follows from standard minimax techniques taking single harmonic eigenfunctions as trial states \cite{GGJL,RS4}. So collapse is actually easier in that case.}
\item satisfy $|\dl V| >> e^{-d_V/\hbar}$ for sufficiently small $\hbar$ (e.g., by being  independent of $\hbar$);
\item\label{co3}  be localized not too far from at least one the minima, in the following sense. 

First, for $y,z\in\R$  and $A\subset\R$, we extend the notation \er{Cwkb} to 
\begin{eqnarray}
d_V(y,z)&=&\left| \int_y^z dx\, \sqrt{V(x)}\right|;\label{Cwkb2}\\
d_V(y,A)&=& \inf\{ d_V(y,z), z\in A\}.
\end{eqnarray}
Second, we introduce the symbols
\begin{eqnarray}
d_V'&=&2\cdot \min\{d_V(-a,\mbox{supp}\ \dl V),d_V(a,\mbox{supp}\ \dl V)\};\\
d_V''&=&2\cdot \max\{d_V(-a,\mbox{supp}\ \dl V),d_V(a,\mbox{supp}\ \dl V)\}.
\end{eqnarray}
The localization assumption on $\dl V$ is that one of the following conditions holds:
\begin{eqnarray}
d_V'< d_V< d_V'' ; && \label{case1a} \\
d_V'< d_V''< d_V \label{case2a}.
\end{eqnarray}
In the first case, the perturbation is typically localized either on the left or on the right edge of the double well, whereas in the second it resides
 on the middle bump.\footnote{Symmetric perturbations are excluded by \ref{co3}, as these would satisfy  $d_V'=d_V''$.}
\end{enumerate}
Under these assumptions,  the ground state wave-function $\Ps^{(\dl)}_{\hbar}$ of the perturbed Hamiltonian (which had two peaks for $\dl V=0$!) localizes as $\hbar\raw 0$, in a direction which \emph{given that localization happens} may be understood from energetic considerations. For example,
if $\dl V$ is positive and is localized to the right, then the relative energy in the left-hand part of the double well is lowered, so that localization will be to the left. See Figures \ref{Fleap1} - \ref{Fleap3}.\newpage
 \vspace*{-5mm}
\vspace*{-5mm}

\noindent In more detail, for the perturbed ground state we have (subject to assumptions 1--4): 
\begin{eqnarray}
{\frac{\Ps^{(\dl)}_{\hbar}(a)}{\Ps^{(\dl)}_{\hbar}(-a)}}&\sim&  e^{\mp d_V/\hbar} \:\:\: (\pm \dl V>0, \mbox{ supp} (V)\subset \R^+);
\label{case1}\\
{\frac{\Ps^{(\dl)}_{\hbar}(a)}{\Ps^{(\dl)}_{\hbar}(-a)}}&\sim&  e^{\pm d_V/\hbar} \:\:\: (\pm \dl V>0, \mbox{ supp} (V)\subset \R^-),\label{case3}
\end{eqnarray}
with the opposite localization for the perturbed first excited state (so as to remain orthogonal to the ground state).\footnote{If $\dl V$ has support on both sides of the real axis (which is possible in the case \er{case2a}), a more detailed analysis of its shape  is necessary in order to predict the direction of collapse \cite{Cesi1}.}  A more precise version of the energetics used above is then as follows. The ground state tries to minimize its energy according to the rules:\footnote{Compare \cite[p.\ 35]{RS4,Simon4} for such arguments. Nonetheless, the effect of the flea is counterintuitive even from the point of view of quantum-mechanical tunneling: for example, with a perturbation of the kind displayed in Figures  \ref{Fleap1} - \ref{Fleap3}, which falls under case \er{case2a}, one would expect tunneling from the right into the left-handed well to be discouraged, even increasingly so as $\hbar\raw 0$, because the potential barrier through which to tunnel has been heightened,
 but in fact the right-handed peak of the unperturbed ground state tunnels to the left so as to localize the ground state wave-function. See  \S  \ref{secQM} for further discussion.}
\begin{itemize}
\item The cost of localization (if $\dl V=0$)  is $\mathcal{O}(e^{-d_V/\hbar})$.
\item The cost of turning on $\dl V$ is $\mathcal{O}(e^{-d_V'/\hbar})$ when the wave-function is delocalized.
\item The cost of turning on $\dl V$ is $\mathcal{O}(e^{-d_V''/\hbar})$ when the wave-function is localized in the well around $x_0=\pm a$ for which $d_V(x_0,\mbox{supp}\ \dl V)=d_V''$.
\end{itemize}
In any case, these results only depend on the support of $\dl V$,  but not on its size: this means that the tiniest of perturbations may cause collapse in the classical limit. 

Although the collapse of the perturbed ground state for small $\hbar$  is a mathematical theorem, supported (or rather illustrated) both by our numerical simulations and by the {\sc wkb} analysis in the appendix, it remains an enigmatic phenomenon of a purely quantum-mechanical nature. 
Indeed, despite the fact that in quantum theory the localizing  effect of the flea is enhanced for small $\hbar$, the corresponding classical system has no analogue of it. Trivially, a classical particle residing at one of the two minima of the double well at zero (or small) velocity, i.e., in one of its degenerate ground states, will not even notice the flea; the ground states are unchanged. But even under a stochastic perturbation, which leads to a nonzero probability for the particle to be driven from one ground state to the other in finite time (as some form of classical ``tunneling'', where in this case the necessary fluctuations come from Brownian motion), the flea plays a negligible role. For example,  in the case at hand  the famous Eyring--Kramers formula for the mean transition time reads
\begin{equation}
\langle \ta\rangle \cong \frac{2\pi}{\sqrt{V''(a)V''(0)}}e^{V(0)/\epsilon},
\end{equation}
where $\ep$ is the parameter in the pertinent Langevin equation $dx_t=-\nabla V(x_t)dt +\sqrt{2\ep} dW_t$, in which $W_t$ is standard Brownian motion.\footnote{Cf.\ \cite{Berglund} (for mathematicians)  or \cite{HTB} (for physicists), and references therein.} Clearly, this expression  only contains the height of the potential at its maximum and its curvature at its critical points; most perturbations satisfying assumptions 1--4 above do not affect these quantities. 
\subsection{Two-level approximation}
The instability of the ground state  of the double-well potential under ``flea'' perturbations as $\hbar\raw 0$ is easy to understand (at least heuristically) if one truncates 
the infinite-dimensional \Hs\ $L^2(\R)$ to a two-level system living in $\C^2$
 \cite{Helffer,Simon4}.\footnote{This approximation is extremely well known also in physics \cite{Leggett}, but has hardly been studied in the present context. It is too simple to display the behaviour \er{solcat} - \er{solcat3}, though.
See also \S\ref{univ}. }
 This simplification is accomplished by keeping only
 the lowest energy states $\Ps_{\hbar}^{(0)}$ and $\Ps_{\hbar}^{(1)}$,  in which case
 the full Hamiltonian   \er{TheHam}  with \er{Vq3}  is reduced to the $2\x 2$ matrix
\begin{equation}
H_0=\half
\left(
\begin{array}{cc}
 0 & -\Dl     \\
 -\Dl &  0
 \end{array}
\right), \label{HD}
\end{equation}
with $\Dl>0$ given by \er{ED}. We drop the label $\hbar$. 
The eigenstates of $H_0$ are given by
\begin{equation}
\Phi_0^{(0)}=\frac{1}{\sqrt{2}}\left(
\begin{array}{c}1\\ 1 \end{array}\right), \:\:\: \Phi_0^{(1)}=\frac{1}{\sqrt{2}}\left(
\begin{array}{c}1\\ -1 \end{array}\right), \label{double}
\end{equation}
with energies $E_0=-\half\Delta$ and $E_1=\half\Dl$, respectively; in particular, $E_1-E_0=\Dl$. If 
\begin{equation}
\Phi_0^{\pm}= \frac{\Ph_0^{(0)}\pm\Ph_0^{(1)}}{\sqrt{2}}, \label{plusmin2}
\end{equation}
 as in \er{plusmin}, then 
\begin{equation}
\Phi_0^+=\left(
\begin{array}{c}0\\ 1 \end{array}\right), \:\:\: \Phi_0^-=\left(
\begin{array}{c}1\\ 0  \end{array}\right). \label{double2}
\end{equation}
Hence in this approximation $\Ph_0^+$ and $\Phi_0^-$ play the role of wave-functions \er{plusmin}  localized above the classical minima $x=+a$ and $x=-a$, respectively, with classical limits $\rh_0^{\pm}$. The ``flea'', then,  is introduced as follows: if its support is in $\R^+$, then we put
\begin{equation}
\dl_+ V=\left(
\begin{array}{cc}
0 & 0\\
 0 &  \dl
 \end{array}
\right), \label{PR}
\end{equation}
where $\dl\in\R$ is a constant, whereas a perturbation with support in $\R^-$ is approximated by
\begin{equation}
\dl_- V=\left(
\begin{array}{cc}
\dl & 0\\
 0 &   0
 \end{array}
\right).\label{PL}
\end{equation}
Without loss of generality, let us take the latter (a change of sign of $\dl$ leads to the former). The eigenvalues of $H^{(\dl)}=H_0+\dl_- V$ are
$E_0=E_-$ and $E_1=E_+$, with energies
\begin{equation}
E_{\pm}= \half(\dl\pm\sqrt{\dl^2+\Dl^2}), \label{Eplm}
\end{equation}
and normalized eigenvectors
\begin{eqnarray}
\Phi^{(0)}_{\dl} &=&\frac{1}{\sqrt{2}}\left(\dl^2+\Dl^2 +\dl\sqrt{\dl^2+\Dl^2}\right)^{-1/2}\left( \begin{array}{c}
\Dl \\
\dl+\sqrt{\dl^2+\Dl^2} \end{array} \right);\\
\Phi^{(1)}_{\dl} &= &
\frac{1}{\sqrt{2}}\left(\dl^2+\Dl^2 -\dl\sqrt{\dl^2+\Dl^2}\right)^{-1/2}\left( \begin{array}{c}
\Dl \\
\dl-\sqrt{\dl^2+\Dl^2} \end{array} \right).
\end{eqnarray}
Note that $\lim_{\dl\raw 0}\Phi^{(i)}_{\dl}=\Phi^{(i)}_0$ for $i=0,1$. 
Now, if $\hbar\raw 0$, then $|\dl|>>\Dl$, in which case  $\Ph^{(0)}_{\dl}\raw \Phi_0^{\pm}$ for  $\pm \dl>0$
(and if we had started from \er{PR} instead of \er{PL}, one would have had the opposite case, i.e., $\Ph^{(0)}_{\dl}\raw \Phi_0^{\mp}$ for  $\pm \dl>0$).
Thus the ground state localizes as $\hbar\raw 0$, which  resembles the situation \er{case1} - \er{case3} for the full double-well problem. 
\section{Time-dependent collapse}\label{Dynamical}
As remarked in Section \ref{sec2.3}, 
for a solution of the \mmp\ it is not enough to just note that under a typical ``flea'' type perturbation (as defined in Section \ref{sec3.1}) the ground state  $\Ps_{\hbar}^{(\dl)}$  of the perturbed Hamiltonian is localized. In addition, the archetypal Schr\"{o}dinger Cat state $\Ps_{\hbar}^{(0)}$, which results from some measurement, needs to evolve into $\Ps_{\hbar}^{(\dl)}$ under the influence of this perturbation. This is a problem in quantum metastability. As a first orientation, we continue our discussion of the two-level system. 
The simplest idea would be to launch the flea as a so-called ``quench'', which means that for times $t<0$ the dynamics is given by $H_0$, upon which for $t\geq 0$ the Hamiltonian is  $H_0+\dl_-V$. Hence
\begin{equation}
H(t)=
\left(
\begin{array}{cc}
 \dl(t) & -\half\Dl     \\
 -\half \Dl &  0
 \end{array}
\right),\label{HD2}
\end{equation}
where $\dl(t)=0$ for $t< 0$ and $\dl(t)=\dl$ for $t\geq 0$. Writing $\Phi^{(0)}(t)$ for the solution of the corresponding time-dependent  Schr\"{o}dinger equation with initial condition $\Phi^{(0)}(0)=\Phi^{(0)}_0$, see \er{double}, for the localization probability ``on the left'', i.e., above $x=-a$, we  find
\begin{equation}
P_L(t)\equiv |\la \Phi_0^-, \Phi^{(0)}(t)\ra|^2=P_L(0)+\half\frac{\dl\Dl}{\dl^2+\Dl^2}\cdot\left[\cos\left(\frac{it}{\hbar}\sqrt{\dl^2+\Dl^2}\right)-1\right],
\end{equation}
where $\Phi_0^-$ is given in \er{double2}. Since $\dl\Dl/(\dl^2+\Dl^2)\raw 0$ as $\hbar\raw 0$, we see from this and similar calculations for other initial states that for any $t$ (including $t\raw\infty$ in whatever, even  $\hbar$-dependent, way),  in the classical limit the initial state freezes rather than collapses. 
\smallskip

\begin{figure}[H]
\begin{center}
\includegraphics[width=0.32\textwidth]{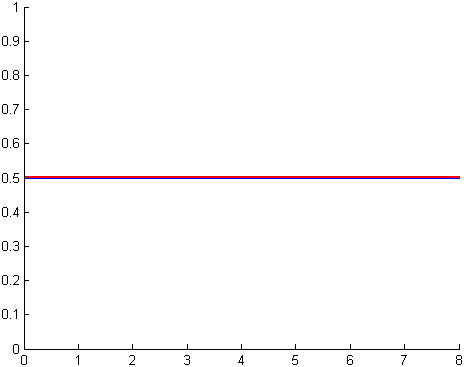}
\includegraphics[width=0.32\textwidth]{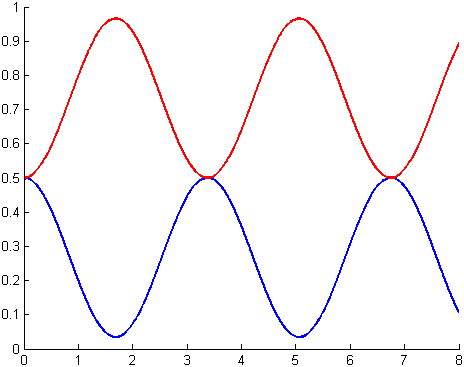}
\includegraphics[width=0.32\textwidth]{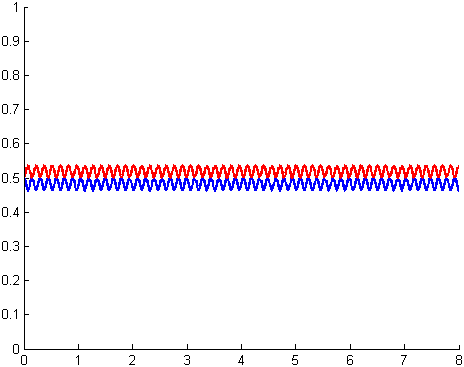}
\caption{Time evolution of the probabilities \blue{$P_L(t)$} and \red{$P_R(t)=1-P_L(t)$}.
The left image has $\dl=0$,  the middle has moderate $\delta$,  and the third  (displaying ``freezing'') has large $\dl$.}
\label{TLSConstant}
\end{center}
\end{figure}
\vspace{-5mm}
Towards less naive time-dependent models for the flea perturbation, we also investigated adding white noise or Poisson noise to the  time-dependent  Schr\"{o}dinger equation. In the two-level case, the pertinent
stochastic differential equations are
\begin{eqnarray}
d\Phi&=& -(\half i\Dl \sg_x dt+i\dl\sg_z dB_t+\half\dl^2 dt)\Phi;\\
d\Phi&=&-(\half i\Dl \sg_x dt+ (\sg_z-\mathbb{I}_2)dN_t)\Phi,
\end{eqnarray}
respectively, where $(\sg_k)$ are the Pauli matrices, $B_t$ is Brownian motion, and $N_t$ is a Poisson process, both with tunable parameters. 
However, neither of these leads to dynamical collapse in the classical limit: this is equivalent to strong noise, in which case  a quantum Zeno-like effect seems to dominate any desire of the system to localize.
See also  \cite{Berry, Blanchard1, MS2}.
Similar (negative) conclusions follow for the full double well (at least, numerically).
\medskip

As far as we have been able to determine, the most effective way to produce dynamical collapse is to let the flea jump on the cat adiabatically (cf.\  \cite{Griffiths}). This is easily shown for the two-level system, but we might as well return to the full double-well problem here. We perturb this potential $V$ with a flea with 
center $b$, width $2c$, and height $d$, as follows:\begin{equation}
\dl V_ {b,c,d}(x)=\left\{\begin{array}{ll}
d\cdot e^{\frac{1}{c^2}-\frac{1}{c^2-(x-b)^2}}& \mbox{if \ } |x-b|<c\\
0 & \mbox{if \ } |x-b|>c
\end{array}\right. .
\end{equation}
One attractive possibility is then to let this perturbation arises adiabatically according to
\begin{equation}
V(x,t)=\left\{\begin{array}{ll}
V(x) & \mbox{if \ } t\leq 0;\\
V(x)+\dl V_{b,c,d}(x)\sin\left(\frac{\pi t}{2T}\right) & \mbox{if \ } 0\leq t\leq T;\\
V(x)+\dl V_{b,c,d}(x) & \mbox{if \ } t>T.
\end{array}\right. 
\label{RisingPot}
\end{equation}
The time-dependent Schr\"{o}dinger equation can be solved numerically with the ground state $\Ps_{\hbar}^{(0)}$ as the initial condition at $t=0$. This yields the following pictures,\footnote{A corresponding movie may be found on \texttt{www.math.ru.nl/$\sim$landsman/flea.avi}.} in which dynamical localization is clearly visible, in agreement with the adiabatic theorem.\footnote{This states that
under an adiabatic perturbation $\dl V=\dl V(t/T)$ the unperturbed ground state moving under the perturbed Hamiltonian converges to the ground state of the latter as $t\raw T\raw\infty$ \cite{Griffiths,HJ}.}
 \begin{figure}[H]
\begin{center}
\includegraphics[width=0.49\textwidth]{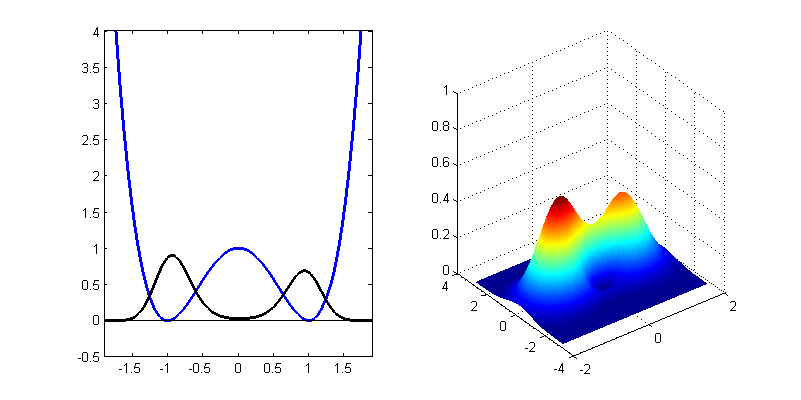}
\includegraphics[width=0.49\textwidth]{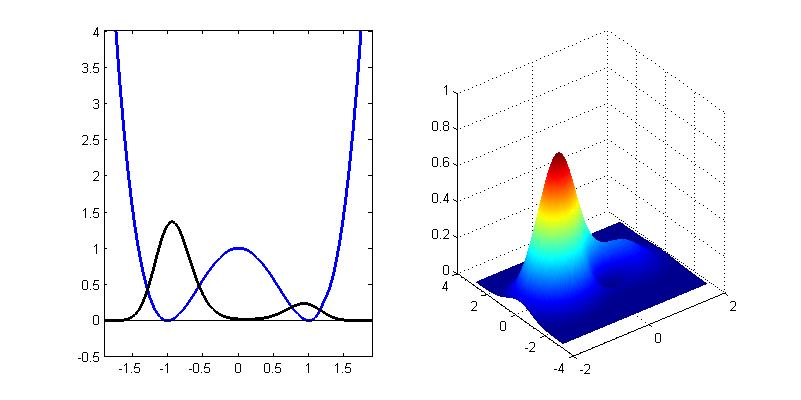}
\bigskip

\includegraphics[width=0.49\textwidth]{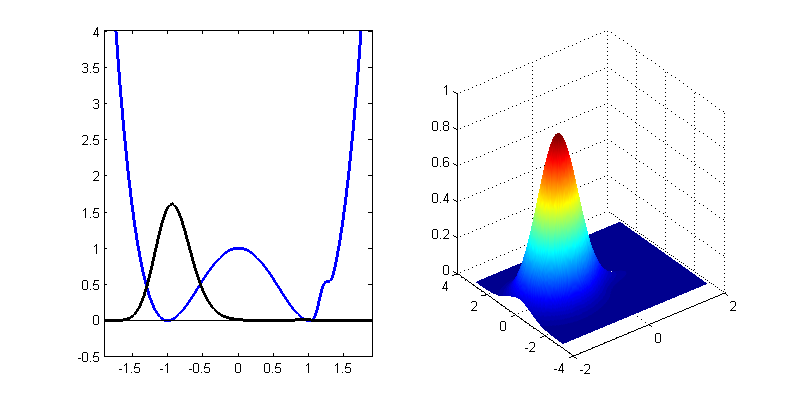}
\includegraphics[width=0.49\textwidth]{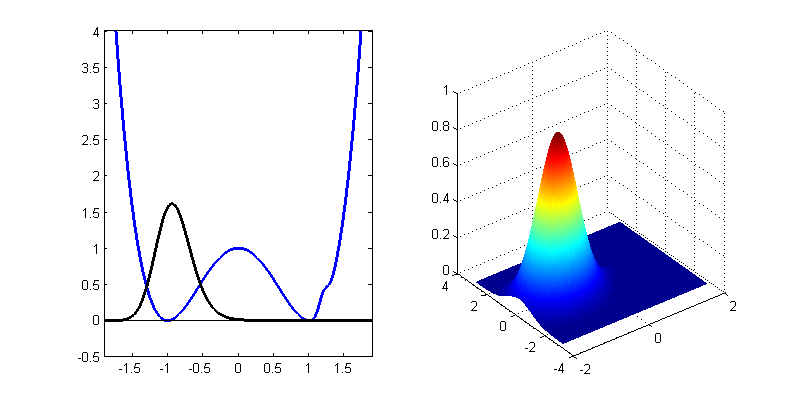}
\caption{Plots of both $|\Ps(t)|^2$ and the corresponding Husumi function for the solution $\Ps(t)$ of the  time-dependent Schr\"{o}dinger equation defined by the potential \er{RisingPot}, with $b=7.5,\ c=0.5,\ d=0.3$, $\hbar=0.3$, and $T=800$. Starting in the upper left corner and proceeding clockwise, the pictures correspond to $t=50$, $t=100$, $t=400$, $t=800$. }
\end{center}
\end{figure}
\vspace{-5mm}
Of course, the symmetry of the situation implies that the Born rule holds if one averages over all perturbations (which corresponds to averaging over a series of experiments) in any reasonable way, i.e.,
any way  in which $\dl V_{b,c,d}$, $\dl V_{-b,c,d}$, $\dl V_{b,c,-d}$, and $\dl V_{-b,c,-d}$ have equal probability. For in that case, according to the rules in Section \ref{sec3.1} any collapse to the left (in some experiment in a long run) will be accompanied by a collapse to the right (in another experiment of the same run) if one of the signs changes. Hence the probabilities for collapse to the left and to the right will both equal $1/2$, in agreement with \er{1.6}.
\newpage 
\section{Discussion}
In this section we deal with various issues around the flea, mostly calling for further research. This will  put our results in perspective, in particular addressing the question to what extent an appropriate generalization of the mechanism we propose may be expected to eventually solve the \mmp. 
At this early stage, we just claim that our simple example holds at least the promise of  generalization, in our opinion even to such an extent that the underlying mechanism for collapse, special and ad hoc as it may undoubtedly look at first sight,  may be universal. 
We explain our optimism in \S\ref{univ} and \S\ref{SBPT}. In \S\ref{secQM} we draw attention to the main dynamical issue behind our proposal, and finally in 
\S\ref{Bell} we briefly discuss the role of determinism, locality, and Bell's Theorem. 
\subsection{Universality of the mechanism}\label{univ}
If anything, a post-measurement state must be stable. This suggests that ground states are natural candidates. Furthermore, $n$ distinct ``peaks'' of the wave-function of such a ground state should bijectively  correspond to $n$ distinct measurement outcomes. This can be realized, for example, in the double-well potential in the classical limit, as discussed in the present paper, but also in the setting of large systems in the thermodynamic limit (cf.\  \S\ref{SBPT} below). Either way, the abstract mechanism behind the collapse of the wave-function is that the system has $n$ low-lying energy eigenstates (of which the ground state is just one), which become degenerate in the appropriate (i.e. classical or thermodynamic) limit, and whose energy levels 
do not cross the remainder of the spectrum in this limit.
Localized perturbations that hardly change these energy levels  and hence preserve this separation then  suffice to sufficiently shake up the ground state  that it collapses to one of these peaks. 

Beyond calculations and simulations like the ones presented in the main body of the present paper, this effect is best studied abstractly, using the so-called \emph{interaction matrix} introduced by Helffer and Sj\"{o}strand \cite{Helffer,HS,Simon4}. 
This matrix acts on an $n$-dimensional vector space $V$ spanned by a basis of conveniently chosen 
localized wave-functions,\footnote{For potentials with $n$ identical wells, this basis may be chosen to consist of the eigenfunctions of the $n$ individual wells with Dirichlet boundary conditions  \cite{Helffer,HS}. For the quantum Ising model (cf.\ \cite{KT}), where $n=2$, one may take the state with all spins up and the state with all spins down, etc.}  whose linear span (i.e.\ $V$) exponentially well approximates the linear span of the first $n$
 exact eigenfunctions of  the unperturbed Hamiltonian $H_{\hbar}$ (like  \er{TheHam}), 
 as $\hbar\raw 0$. 
In this basis, the ground state of the  $H_{\hbar}$ is approximately $(1,1,\ldots,1)/\sqrt{n}$, displaying the $n$ peaks of the post-measurement state. 
Now if a perturbation $\dl V$ has the properties just mentioned, then the lowest $n$ levels and eigenstates of $H+\dl V$ can be studied using this compressed form, resulting in the collapse studied throughout this paper: typically,  as $\hbar\raw 0$  the (approximate) perturbed ground state
is one of the basis vectors itself  and hence is localized. 

This mechanism only uses linear algebra, and hence it is independent of the use of the position representation in our basic double well example.\footnote{ In this respect our approach has an edge over Bohmian mechanics and the {\sc grw} theory, although it remains to be shown that typical environments induce the right perturbations for the mechanism to apply.} For example, in large spin systems (see below) the appropriate notion of localization refers to spin configurations rather than position. All that matters is that
the ground state of the original,  unperturbed Hamiltonian (which arguably models a typical post-measurement state) has $n$ peaks that become well separated in the appropriate (classical or thermodynamic) limit, with $n$ corresponding energy levels not crossing the rest of the spectrum in the limit.\subsection{Symmetry breaking and phase transitions}\label{SBPT}
The symmetric double-well potential  provides one of the simplest models of spontaneous symmetry breaking ({\sc ssb}). Both the classical Hamiltonian  and 
 its quantization  have a $\Z_2$ symmetry, given by reflection in the origin of the $x$-axis. 
As we have seen, a remarkable difference between classical and \qm\ arises:
 the classical ground state is degenerate and breaks the symmetry, whereas the quantum ground state $\Ps^{(0)}_{\hbar}$ is unique and hence symmetric. If we see the splitting of the ground state as a phase transition,  then evidently the quantum system has no phase transition, whereas its classical counterpart does. At first sight this appears to be quite paradoxical, since the presence or absence of symmetry breaking is a major \emph{qualitative} feature of the system described either classically or quantum-mechanically, while at the same time we \emph{quantitatively} expect the classical theory to be a limiting case of the quantum theory. Indeed, this is nothing but the \mmp\ in disguise: if, for any $\hbar>0$, the (delocalized) quantum ground state prevails, then the classical ground states $\rh_0^{\pm}$ totally fail to be approximated by it.  
 
 We resolved this problem by the ``flea'' instability. 
 Similarly, the ground state of a large but finite quantum system ($N<\infty$) is typically unique and hence symmetric. But at $N=\infty$,  for suitable Hamiltonians  {\sc ssb} occurs, in that the ground state (or thermal equilibrium state at low temperature) fails to be symmetric. Thus the limit
$N$ \emph{tends to} infinity does not approximate the phenomenon of {\sc ssb} when $N$ \emph{equals} infinity.\footnote{
Like the \mmp, this seemingly paradoxical situation does not seem to bother physicists very much, although their Higgs mechanism relies on a resolution of it: apparently, in any finite volume the system refuses to choose a ground state (or vacuum), although all perturbative calculations underlying the successful Standard Model of elementary particle physics rely on such a choice. 
 But it has been the subject of recent discussions in the philosophy of science \cite{Bat,Butt,LiuEmch,MC,Norton}, in which some claim that this ``discontinuity'' in passing from $N<\infty$ to $N=\infty$ is crucial for the possibility of emergence (`More is Different'), whilst others try to find arguments for continuity and hence defend some form of reductionism.}
 Based on \cite{KT} and the discussion in \S\ref{univ}, we expect to find some analogue of the ``flea'' perturbation and expect it to be especially effective for large $N$. This should destabilize the ground state so as to break the symmetry already in large \emph{but finite} volume. Indeed, an instability like this \emph{must} underlie the Higgs mechanism (proved phenomenologically relevant in 2012).
  \subsection{Quantum metastability}\label{secQM}
As we see it, the \mmp\ is ultimately a problem in \emph{quantum metastability}. 
 Metastability is well understood if it is thermally driven, both in classical and in quantum theory 
 \cite{HTB, OlVa05,Sewell0,Sewell1}, but in our approach the driving force is neither a heat bath nor Brownian motion: quantum fluctuations are supposed to drive the old (unperturbed) ground state to the new (perturbed) one. Little is known about this situation even heuristically
 \cite{Nieto,Shifman}.  What we can say is that the limit $\hbar\raw 0$ seems a double-edged sword: on the one hand, it causes the instability of the original  ground state, and hence favours \emph{static} collapse (in that the perturbed ground state is localized, unlike the original one), but on the other hand, it suppresses tunneling and hence slows down \emph{dynamical} collapse (i.e.\ the perturbed time-evolution of  the unperturbed ground state into the localized perturbed one). Together with the adiabatic arrival of the flea,\footnote{If it happens to be true that measurement outcomes emerge adiabatically, it would be a marked break with tradition, starting with von Neumann's model, in which both the measurement interaction and the alleged collapse take place instantly \cite{vN32}.  However, the recent notion of a ``weak measurement'' seems to support our approach on this point \cite{Koc,LW,Roz}. We are indebted to Jos Groot for these references. } 
this appears to explain the exceptionally long time it takes for localization to happen in our numerical simulations.   
 \subsection{Determinism and  locality}\label{Bell}
 In so far as determinism is concerned, there are  two ways to look at our proposal.
 
 First, the ``flea'' perturbation might itself be a genuine random process, perhaps being  ultimately of quantum-mechanical origin. In that case, its own intrinsic randomness (whatever that may mean) is simply transferred to the set of possible measurement outcomes. Although in that case the flea may still be said to ``cause'' one particular outcome (of some experiment),  it would fail to restore determinism. Rather, the experiment merely amplifies the randomness that was already inherent in the flea.\footnote{See \cite{ColRen} for a recent discussion of randomness amplification, which focuses on the way experiments may be construed to 
 amplify the randomness inherent in  the (alleged) ``free'' choice of an experimentalist.}
 
Second, the flea might be deterministic (but is just modeled stochastically for pragmatic reasons). This would open the door to a complete restoration of determinism. For now  the flea transfers its \emph{determinism} to the experiment  (rather than its \emph{randomness}, as in the previous scenario). The mistaken impression that quantum theory implies the  irreducible randomness of nature then arises because measurement outcomes are merely
 unpredictable ``for all practical purposes'', indeed in a way that (because of the exponential sensitivity to the flea in $1/\hbar$) dwarfs even the unpredictability of classical chaotic systems. 
 
Either way, the location of the flea plays a similar role to the position variable in Bohmian mechanics \cite{CushingFine,DT}, i.e., it is essentially 
a hidden variable.\footnote{What follows will hardly be new to specialists in these matters, but it needs to be stated clearly.}
 As such, one has to deal with well-known results like Bell's Theorem \cite{Bell1,Bub,Maudlin} or the Free Will Theorem \cite{BS,Clifton,CK1,CK2,HR, Stairs}.
In the analysis of these results, the notions of \emph{Outcome Independence} ({\sc oi}) and \emph{Parameter Independence} ({\sc pi}), have come to play an important role, and these are especially relevant to our situation.\footnote{Using the traditional scene where (Alice, Bob) are spacelike separated and perform experiments with settings $(a,b)$ and outcomes $(x,y)$, respectively, a stochastic \hv\ $\lm$ (or rather the corresponding theory) satisfies {\sc oi} if Alice's conditional probability $P(x|a,b,y,\lm)$ of finding $x$ given $(a,b,y,\lm)$ is independent of Bob's outcome $y$, whereas the theory satisfies  {\sc pi} if her conditional probability $P(x|a,b,\lm)$ of finding $x$ given $(a,b,\lm)$ is independent of Bob's 
setting $b$, and \emph{vice versa}.
These notions  were originally introduced, in slightly different form, by Jarrett \cite{Jarrett}. Controversies around this terminology and its use of the sort discussed in e.g.\ \cite{Bub, Ghirardi,Maudlin,Norsen1,Seevinck} seem irrelevant to our purposes.} Briefly, the conjunction of {\sc oi} and {\sc pi} is equivalent to Bell's locality condition, and if the latter is satisfied, then also the Bell inequalities hold. Since these are violated by \qm, any \hv\ theory compatible with \qm\ must violate either {\sc oi} or  {\sc pi} (or both). Deterministic \hv\ theories necessarily satisfy {\sc oi}, in which case both Bell's Theorem and the  Free Will Theorem show that they must violate {\sc pi} in order to be compatible with \qm. 
A violation of {\sc pi} leads to possible superluminal signalling only if the variable $\lm$ can be controlled \cite{Maudlin,Norsen1}. 
If the wave-function $\Ps$ is regarded as the ``hidden'' (sic) variable $\lm$, then quantum theory itself satisfies {\sc pi} but violates {\sc oi} (since $\Ps$ \emph{can} be controlled, the other way round would be disastrous). Being a deterministic \hv\ theory, Bohmian mechanics satisfies  {\sc oi}, and hence it violates {\sc pi} \cite{Maudlin}. The {\sc grw} theory, on the other hand, satisfies  {\sc pi} but violates {\sc oi} \cite{BG2,Ghirardi,GBF,Norsen1}. 

The fate of our own approach depends on the nature of the perturbation: if the flea is deterministic, our theory behaves like Bohmian mechanics in this respect  and hence violates {\sc pi}, whereas stochastic perturbations  typically violate {\sc oi} (and possibly also {\sc pi}). 

Either way, no conflict with the said theorems arises. Paraphrasing a comment often made by Bell concerning Bohmian mechanics and {\sc grw} \cite{Maudlin}: the nonlocality of the collapse mechanism we propose just reflects the nonlocality inherent in \qm\ itself. 
\section{Appendix: the flea from  {\sc wkb}}
In this appendix, we study  the ``flea'' type perturbation from the point of view of
 the {\sc wkb} method of the physics textbooks (like \cite{Griffiths,LL}).\footnote{As opposed to the extremely sophisticated and mathematically rigorous methods of Helffer and Sj\"{o}strand \cite{DS,Helffer,HS}, who somewhat confusingly suggest they use the ordinary  {\sc wkb} method.} As explained in \cite{Froman2,Froman}, the  connection formulae stated in such books are actually correct only for simple potentials like a single well, but 
 with due modifications (see below), the formalism  will reproduce both the rigorous and the numerical results described in the main body of this paper.
\subsection{Quantization condition for an asymmetric double well}
\begin{figure}[H]
\begin{center}
\includegraphics[width=7.5cm]{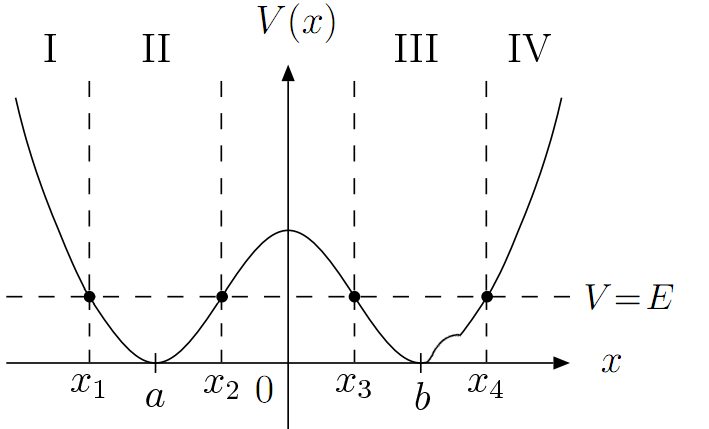}
\caption{An asymmetric double-well potential $V$. The minima are $a$ and $b$.
We assume that the particle has energy $E$. This provides us with turning points $x_1$, $x_2$, $x_3$ and $x_4$ and hence with five distinct regions.
Four of these regions are named with Roman numerals.}
\label{doublewell2}
\end{center}
\end{figure}

 We start by recalling some standard {\sc wkb} formulas. 
The {\sc wkb} wave-function in the classically allowed region without turning points ($E>V(x)$) can be written as
\begin{equation}
\Ps(x)\cong\frac{1}{\sqrt{p(x)}}\left[Ae^{\frac{i}{\hbar}\int^{x}{p(y)dy}}+Be^{-\frac{i}{\hbar}\int^{x}{p(y)dy}}\right],
\label{WKBclassical}
\end{equation}
where
\begin{equation}
p(x)=\left\{\begin{array}{ll}
\sqrt{\left[E-V(x)\right]} & \mbox{if } E\geq V(x)\\
\pm i\sqrt{\left[V(x)-E\right]} & \mbox{if } E< V(x)
\end{array}\right. .
\label{p}
\end{equation}
 A similar formula holds for the classically forbidden region ($E<V(x)$), namely
\begin{equation}
\Ps(x)\cong\frac{1}{\sqrt{|p(x)|}}\left[Ce^{-\frac{1}{\hbar}\int^{x}{|p(y)|dy}}+De^{\frac{1}{\hbar}\int^{x}{|p(y)|dy}}\right].
\label{WKBforbidden}
\end{equation}
 These wave-functions can be connected across turning points via so-called connection formulas, stated in books like \cite{Griffiths}.
First, we need to distinguish between two kinds of turning points in the usual way: we use the coefficients $A_l$, $B_l$, $C_l$ and $D_l$ for a left-hand turning point and $A_r$, $B_r$, $C_r$ and $D_r$ for a right-hand one. 
The lower limit of the integrals in the above equations is always the coordinate of the turning point.
The connection formulas for a left-hand turning point are given by
\begin{equation}
\left( \begin{array}{c}
A_l\\
B_l\\
\end{array} \right)=\overbrace{e^{i\pi/4}
\left( \begin{array}{cc}
\half & -i\\
-\frac{i}{2} & 1\\
\end{array} \right)}^{M_{C_l/D_l\rightarrow A_l/B_l}}
\left( \begin{array}{c}
C_l\\
D_l\\
\end{array} \right)
\mbox{ or }
\left( \begin{array}{c}
C_l\\
D_l\\
\end{array} \right)=\overbrace{e^{-i\pi/4}
\left( \begin{array}{cc}
1 & i\\
\frac{i}{2} & \half\\
\end{array} \right)}^{M_{A_l/B_l\rightarrow C_l/D_l}}
\left( \begin{array}{c}
A_l\\
B_l\\
\end{array} \right),
\label{connection1}
\end{equation}
whilst  those for a right-hand turning point are given by
\begin{equation}
\left( \begin{array}{c}
A_r\\
B_r\\
\end{array} \right)=
\overbrace{e^{i\pi/4}\left( \begin{array}{cc}
1 & -\frac{i}{2}\\
-i & \half\\
\end{array} \right)}^{M_{C_r/D_r\rightarrow A_r/B_r}}
\left( \begin{array}{c}
C_r\\
D_r\\\end{array} \right)
\mbox{ or }
\left( \begin{array}{c}
C_r\\
D_r\\
\end{array} \right)=\overbrace{e^{-i\pi/4}
\left( \begin{array}{cc}
\half & \frac{i}{2}\\
i & 1\\
\end{array} \right)}^{M_{A_r/B_r\rightarrow C_r/D_r}}
\left( \begin{array}{c}
A_r\\
B_r\\
\end{array} \right).
\label{connection2}
\end{equation}
Now consider a general asymmetric double well, as shown in Figure \ref{doublewell2}. This figure also introduces part of the notation used.
We need some more notation for the {\sc wkb} coefficients used in our calculation.
As in \eqref{WKBclassical} and \eqref{WKBforbidden}, $A,B$ and $C,D$ denote the coefficients of the {\sc wkb} wave-function in the classically allowed region and the classically forbidden region, respectively.
The number attached to a letter shows to which turning point it belongs, e.g.\ $A_1$ and $B_1$ are the coefficients of the {\sc wkb} wave-function in region II with respect to $x_1$ (i.e.\ $x_1$ is the lower boundary of the integral in \eqref{WKBclassical}).
We also need the following three quantities:
\begin{equation}
\theta_1=\frac{1}{\hbar}\int^{x_2}_{x_1}p(x)dx, \hspace{1cm} \theta_2=\frac{1}{\hbar}\int^{x_4}_{x_3}p(x)dx,
\hspace{1cm} K=\frac{1}{\hbar}\int^{x_3}_{x_2}|p(x)|dx.
\end{equation}
A final quantity we need is
\begin{equation}
\tilde{\phi}=\arg{\left[\Gamma\left(\half+i\frac{K}{\pi}\right)\right]}+\frac{K}{\pi}-\frac{K}{\pi}\ln\left(\frac{K}{\pi}\right).
\label{constantsWKB}
\end{equation}
We are interested in the limit $K\rightarrow\infty$, since this implies that the barrier is very high and broad, which corresponds to the classical limit $\hbar\rightarrow0$.
Note that $\tilde{\phi}\rightarrow 0$ as $K\rightarrow\infty$.
 Our goal is the following quantization condition for the general double well in Figure \ref{doublewell2}:
\begin{equation}
\left(1+e^{-2K}\right)^{1/2}=\frac{\cos(\theta_1-\theta_2)}{\cos(\theta_1+\theta_2-\pi+\tilde{\phi})}\ .
\label{AsymetricQuantization}
\end{equation}
This condition can be derived in the following way:
\begin{enumerate}
\item We start out in region I (coefficients $C_1$ and $D_1$).
The wave-function needs to be square integrable, so we immediately see that $C_1=0$.
\item Using the left connection matrix from \eqref{connection1}, we move to region II (coefficients $A_1$ and $B_1$).
We can then write the {\sc wkb} wave-function with respect to $x_2$ by using
\begin{equation}
\left( \begin{array}{c}
A_2\\
B_2\\
\end{array} \right)=
\left( \begin{array}{cc}
e^{i\theta_1} & 0\\
0 & e^{-i\theta_1}\\
\end{array} \right)
\left( \begin{array}{c}
A_1\\
B_1\\\end{array} \right),
\end{equation}
which can be proved by changing the lower boundary of the integrals in the {\sc wkb} wave-function \eqref{WKBclassical}.
The result is 
\begin{equation}
\left( \begin{array}{c}
A_2\\
B_2\\
\end{array} \right)=e^{i\pi/4}
\left( \begin{array}{c}
-i e^{i\theta_1}\\
\ \ \ e^{-i\theta_1}\\
\end{array} \right)D_1.
\label{aaa}
\end{equation}
\item In a similar way, we start in region IV (coefficients $C_4$ and $D_4$), and see that $D_4=0$.
After moving to region III with a connection matrix and rewriting the wave-function with respect to $x_3$, we find
\begin{equation}
\left( \begin{array}{c}
A_3\\
B_3\\
\end{array} \right)=e^{i\pi/4}
\left( \begin{array}{c}
\ \ \ e^{-i\theta_2}\\
-i e^{i\theta_2}\\
\end{array} \right)C_4\ .
\label{bbb}
\end{equation}
\item We now use a result derived in \cite{Froman2} to jump over the barrier and connect the {\sc wkb} wave-functions in region II and III, viz.\footnote{This result can also be found by applying the method of comparison equations, which is explained in \cite{Zauderer}.
Further references are \cite{LynnKeller} and \cite{MillerGood}.}
\begin{equation}
\left( \begin{array}{c}
A_2\\
B_2\\
\end{array} \right)=
\left( \begin{array}{cc}
\left(1+e^{2K}\right)^{1/2}e^{-i\tilde{\phi}} & ie^K\\
-ie^K & \left(1+e^{2K}\right)^{1/2}e^{i\tilde{\phi}}\\
\end{array} \right)
\left( \begin{array}{c}
A_3\\
B_3\\\end{array} \right).
\label{ccc}
\end{equation}
\item Combining the above results (i.e.\ inserting \eqref{aaa} and \eqref{bbb} in \eqref{ccc}), we find 
\begin{align}
\label{WKBLocalization}
\frac{D_1}{C_4} &=i\left[\left(1+e^{2K}\right)^{1/2}e^{-i(\theta_1+\theta_2+\tilde{\phi})}+e^Ke^{-i(\theta_1-\theta_2)}\right],\\
\frac{D_1}{C_4} &=-i\left[\left(1+e^{2K}\right)^{1/2}e^{i(\theta_1+\theta_2+\tilde{\phi})}+e^Ke^{i(\theta_1-\theta_2)}\right].
\end{align}
\item The equality of the above two equations leads to the quantization condition \eqref{AsymetricQuantization}.
\end{enumerate}
As will be discussed in the next two subsections, eqs.\ \eqref{AsymetricQuantization} and \eqref{WKBLocalization} have implications for the energy levels and the wave-functions in an asymmetric double well.
\subsection{Energy splitting in an asymmetric double-well potential}
\label{EnSplitWKBAsymm}
Assume that for a certain (unperturbed) symmetric double well and given energy $E$, the constants $\theta_1$ and $\theta_2$ equal some value $\theta$.
As in Figure \ref{doublewell2}, we introduce a perturbation in the right-hand well.
For example, by \eqref{constantsWKB}, this means that $\theta=\theta_1>\theta_2$ for a positive perturbation.
We therefore write $\theta_1=\theta$, $\theta_2=\theta-\delta$ with $\delta\in\mathbb{R}$ (e.g.\ $\delta>0$ in Figure \ref{doublewell2}).
The quantization condition \eqref{AsymetricQuantization} then becomes 
\begin{equation}
\left(1+e^{-2K}\right)^{1/2}=\frac{\cos(\delta)}{\cos(2\theta-\delta-\pi+\tilde{\phi})}\ .
\end{equation}
 We can solve for $\theta$, yielding two solutions
\begin{equation}
\theta_{\pm}=(n+\half)\pi +\half\delta -\half\tilde{\phi} \pm \half\arccos\left[\frac{\cos(\delta)}{\left(1+e^{-2K}\right)^{1/2}}\right].
\label{SolutionTheta}
\end{equation}
This resembles the original quantization condition $\theta=(n+\half)\pi$ for a single well, which is derived using connection formulas in \cite{Griffiths}. 
Here, the energy levels have split up in pairs around the original ones (where the minus sign in \er{SolutionTheta} corresponds to the lower energy by \eqref{constantsWKB}).
To see what this means, we will examine this equation for two special cases.
We first set $\delta=0$ and check if this reproduces known results for a symmetric double well:
\begin{equation}
\theta_{\pm}=(n+\half)\pi -\half\tilde{\phi} \pm \half\arccos\left[\frac{1}{\left(1+e^{-2K}\right)^{1/2}}\right].
\end{equation}
Supposing that $K$ is large, this means that
\begin{equation}
\theta_{\pm}\approx(n+\half)\pi \pm \half e^{-K},
\end{equation}
since for $K$ large, $\tilde{\phi}\approx 0$ and $\arccos\left(\frac{1}{\sqrt{1+x^2}}\right)=\arctan{x}\approx x$ for small $x$.
We once again find that the energy levels of the single well have split into two.
As discussed in \cite{Froman}, this leads exactly to the familiar energy splitting for a symmetric double-well potential stated in texts like \cite{LL}.
That means that  our method for general double wells  reproduces known results for a symmetric one.
Now that this has been confirmed, let us look at \eqref{SolutionTheta} in the classical limit $K\rightarrow \infty$.
Solving \eqref{SolutionTheta} for $K\rightarrow \infty$ (and so $\tilde{\phi}\rightarrow 0$) gives
\begin{eqnarray}
\theta_- &=& (n+\half)\pi \mbox {  (lower energy) };\\
\theta_+&=& \delta+(n+\half)\pi  \mbox {  (higher energy). \label{DeltaKLarge} }
\end{eqnarray}
This  differs from the symmetric well, which for $K\rightarrow \infty$ gives a twofold degeneracy for each energy level labeled by $n$. Equation \eqref{DeltaKLarge} can be understood in the following way: in the classical limit, tunneling is suppressed.
Therefore, the particle is localized in one of the wells, where it obeys the familiar quantization condition for a single well.
If it is in the left well, then $\theta_1=(n+\half)\pi=\theta_-$, but if it is in the right well, we have $\theta_2=(n+\half)\pi=\theta_+ -\delta$.
\subsection{Localization in an asymmetric double-well potential}
\label{LocWKB}
Now that we have analyzed the behaviour of the energy splitting, we turn to the {\sc wkb} wave-function.
With the notation used in the previous section, \eqref{WKBLocalization} leads to
\begin{equation}
\frac{D_1}{C_4} =i\left[\left(1+e^{2K}\right)^{1/2}e^{-i(2\theta_{\pm}-\delta+\tilde{\phi})}+e^Ke^{-i\delta}\right].
\end{equation}
Inserting \eqref{SolutionTheta}, the reader can check that for $\delta\in[-\pi,\pi]$ one has
\begin{equation}
\frac{D_1}{C_4} =\sin(\delta)e^{K} \mp \sqrt{\sin^2(\delta)e^{2K}+1}\ .
\label{D1/C4}
\end{equation}

\noindent This allows us to derive localization of the {\sc wkb} wave-function in the classical limit $K\rightarrow\infty$.
As can be seen from \eqref{aaa}, $D_1$ is a measure of the amplitude of the {\sc wkb} wave-function in regions I and II in Figure \ref{doublewell2}.
In a similar way, \eqref{bbb} shows that $C_4$ is a measure of the amplitude  of the {\sc wkb} wave-function in regions III and IV.
Therefore, the fraction $D_1/C_4$  indicates whether  the wave-function is localized, and if so, where.
Doing the same calculation again for $\delta\in[\pi,3\pi]$ gives the above result multiplied by $-1$.
Of course, this can be generalized: for $n\in\mathbb{Z}$ and $\delta\in[(2n-1)\pi,(2n+1)\pi]$, the result \eqref{D1/C4} is correct for $n$ even and should be multiplied by $-1$ for $n$ odd. 
This will not affect our conclusions, as we will see.
We consider some cases and check what \eqref{D1/C4} tells us:
\begin{itemize}
\item For $\delta=0$ (no perturbation), we find that $\frac{D_1}{C_4}=\mp 1$. 
The general double well has pairs of energy levels (labeled by $n$). 
Such a pair consists of a lower and higher lying level, corresponding to $\theta_-$ and $\theta_+$ in \eqref{SolutionTheta}, respectively.
Here, we see that for the lower level $D_1=C_4$, i.e.\ the {\sc wkb} wave-function is even.
However, for the higher level we find $D_1=-C_4$, which means the {\sc wkb} wave-function is odd.
This is a well-known fact and it is nice to see our method reproducing it.
Note that this conclusion is not only independent of $n$, but also of $K$, as expected.
\item For $\delta>0,\delta\notin\left\{k\pi|k\in\mathbb{Z}\right\}$ (which corresponds to a positive perturbation in the right well, e.g.\ the potential in Figure \ref{doublewell2}), we find, in the limit $K\rightarrow\infty$, that:
\begin{equation*}
\frac{D_1}{C_4}\longrightarrow\left\{\begin{array}{ll}
\infty & \mbox{for $\theta_-$ in \eqref{SolutionTheta} (lower energy)}\\
 0 & \mbox{for $\theta_+$ in \eqref{SolutionTheta} (higher energy)}
\end{array}\right. .
\end{equation*}
Hence for low (high) energy, the {\sc wkb} wave-function is localized on the left (right).
\item For $\delta<0,\ \delta\notin\left\{k\pi|k\in\mathbb{Z}\right\}$, i.e., a negative perturbation in the right well, we find
\begin{equation*}
\frac{D_1}{C_4}\longrightarrow\left\{\begin{array}{ll}
0 & \mbox{for $\theta_-$ in \eqref{SolutionTheta} (lower energy)}\\
\infty & \mbox{for $\theta_+$ in \eqref{SolutionTheta} (higher energy)}
\end{array}\right. .
\end{equation*} 
 For the lower (higher) energy, the {\sc wkb} wave-function is localized on the right (left).
\item For $\delta\in\left\{k\pi|k\in\mathbb{Z}\backslash\{0\}\right\}$, something peculiar happens, in that either $\frac{D_1}{C_4}=\pm 1$ or $\frac{D_1}{C_4}=\mp 1$. This implies that no localization takes place.\footnote{This can be explained by level crossing, i.e.\ certain energy levels of the two individual wells coincide.}

\item So far, we have interpreted $\delta$ as the result of a perturbation in the right well.
However, our approach  allows us to interpret a positive perturbation in the right-hand well as a negative one in the left-hand well, and vice versa. Therefore, the above results change places if we put the perturbation in the left-hand well.
\end{itemize}

\noindent Our method produces the results we would expect. 
However, to be precise, the above reasoning needs to be amended as follows.
We have treated $\delta$ as a constant, but in reality it depends on $K$. 
The reason for this is that $K$ affects $\theta_1$ and $\theta_2$, and therefore $\delta=\theta_1-\theta_2$, via the quantization condition. 
Now  consider a fixed energy level (i.e.\ fixed $n$ and fixed sign $\pm$ in \eqref{SolutionTheta}) in a given double-well potential that has a perturbation in one of the wells.
In the limit of completely decoupled wells ($K\rightarrow\infty$), we know this energy level has some fixed limit higher than the minimum of the potential.
As long as the perturbation is below this energy level, we know that $\theta_1-\theta_2\neq 0$ by \eqref{constantsWKB}.
This means that there exists some $K_0$ such that $|\theta_1-\theta_2|\neq0$ for any $K>K_0$.
We may then apply the above reasoning to verify  that our conclusions about localization are still correct.\footnote{
To keep the discussion straightforward, we ignored the special case $\delta\in\left\{k\pi|k\in\mathbb{Z}\backslash\{0\}\right\}$ here.}
  \subsection*{Acknowledgement}
  The authors are indebted to Koen Reijnders for help with this appendix.
  They also wish to thank Jeremy Butterfield, Fred Muller,  and an anonymous referee for their penetrating comments on the first draft of this paper, which have  improved the current (final) version.\footnote{Muller also notes that this paper reveals the `empiricist presuppositions' of the authors. For example, we are not worried about superpositions like $\Ps=\sum_i c_i \Psi_i$ in \S\ref{histo} as such, despite the fact that most quantum-mechanical observables, including the one ($O$) that is being measured, have no value in such a state. Compared to classical physics this is admittedly a curious feature of quantum theory, but it causes neither  inconsistencies within the formalism nor disagreements with observation. Furthermore, it has nothing to do with $\Ps$ being a superposition (which is a basis-dependent statement), and in our view it only leads to a problem (namely the \mmp) if the device whose state $\Ps$ represents is classical (e.g., in being macroscopic). For in that case \qm\ naively (i.e.\ without the flea mechanism) appears to assign
  \emph{numerous} values to $O$, which seems unacceptable (rather than assigning \emph{no} value, which for the stated reasons we can live with both as theorists and as alleged empiricists). }

\end{document}

%% file: preamble.tex
 \newcommand{\beq}{\begin{equation}}
\newcommand{\eeq}{\end{equation}} 
\newcommand{\bea}{\begin{eqnarray}}
\newcommand{\eea}{\end{eqnarray}}

 \newcommand{\qm}{quantum
mechanics}

\newcommand{\Hs}{Hilbert space}

\newcommand{\raw}{\rightarrow}

\newcommand{\ot}{\otimes} 
\newcommand{\la}{\langle} \newcommand{\ra}{\rangle}

\newcommand{\x}{\times} 
 
\newcommand{\Tr}{\mbox{\rm Tr}\,}

 \newcommand{\cci}{C^{\infty}_c}
\newcommand{\half}{\mbox{\footnotesize $\frac{1}{2}$}}

\newcommand{\quar}{\mbox{\footnotesize $\frac{1}{4}$}}

\newcommand{\inv}{^{-1}}

\newcommand{\er}{\eqref}
\newcommand{\al}{\alpha} 
 
\newcommand{\dl}{\delta} \newcommand{\Dl}{\Delta}
\newcommand{\ep}{\epsilon}

\newcommand{\lm}{\lambda} 
\newcommand{\rh}{\rho} \newcommand{\sg}{\sigma}
 \newcommand{\ta}{\tau} 
\newcommand{\Ph}{\Phi} 
 \newcommand{\ps}{\psi} \newcommand{\Ps}{\Psi}
\newcommand{\om}{\omega} \newcommand{\Om}{\Omega}
\newcommand{\C}{{\mathbb C}} 
 
\newcommand{\N}{{\mathbb N}} \newcommand{\R}{{\mathbb R}}
 \newcommand{\Z}{{\mathbb Z}}
  \makeatletter
\newskip\tempskip \def\endproof{{\parfillskip24\p@ plus\@ne
fil\@@par}\tempskip\prevdepth
\ifdim\lastskip=\z@\tempskip\z@\else\vskip-\lastskip
\ifdim\tempskip>4\p@ \tempskip.5\tempskip \else \tempskip\z@\fi\fi
\nobreak\vskip-\baselineskip\vskip-\tempskip\noindent\hbox
to\hsize{\hfill
$\blacksquare$}\par\vskip\tempskip\vskip\abovedisplayskip\@doendpe}
\makeatother \makeatletter
\newskip\tempskip \def\endiproof{{\parfillskip24\p@ plus\@ne
fil\@@par}\tempskip\prevdepth
\ifdim\lastskip=\z@\tempskip\z@\else\vskip-\lastskip
\ifdim\tempskip>4\p@ \tempskip.5\tempskip \else \tempskip\z@\fi\fi
\nobreak\vskip-\baselineskip\vskip-\tempskip\noindent\hbox
to\hsize{\hfill
$\Box$}\par\vskip\tempskip\vskip\abovedisplayskip\@doendpe}
\makeatother 



%% file: Catresubmission.bbl
\begin{thebibliography}{99}
\bibitem{Wallace} D. Wallace, Decoherence and its role in the modern
measurement problem, \texttt{arxiv:1111.2187}.
\end{thebibliography}
